\documentclass[aps,prx,preprint,onecolumn,citeautoscript,footinbib,eqsecnum]{revtex4-2}
\usepackage{amsmath}
\usepackage{amssymb}
\usepackage{braket}
\usepackage{xcolor}
\usepackage{slashed}
\usepackage{ulem} 
\allowdisplaybreaks

\usepackage{graphicx}
\usepackage{subcaption}
\usepackage[colorlinks=true]{hyperref}
\usepackage{svg}

\renewcommand{\ol}{\overline}
\newcommand{\del}{\partial}
\renewcommand{\vec}[1]{\mathbf{#1}}

\hypersetup{
    bookmarks=true,         
    unicode=false,          
    pdftoolbar=true,        
    pdfmenubar=true,        
    pdffitwindow=false,     
    pdfstartview={FitH},    
    pdfsubject={},   
    pdfcreator={},   
    pdfproducer={}, 
    pdfkeywords={} {} {}, 
    pdfnewwindow=true,      
    colorlinks=true,       
    linkcolor=blue, 
    citecolor=blue,        
    filecolor=magenta,      
    urlcolor=blue           
}

\begin{document}
\title{Pathways from a chiral superconductor to a composite Fermi liquid}

\author{Yunchao Zhang}
\affiliation{Department of Physics, Massachusetts Institute of Technology, Cambridge MA 02139-4307,  USA}
\author{Leyna Shackleton}
\affiliation{Department of Physics, Massachusetts Institute of Technology, Cambridge MA 02139-4307,  USA}
\author{T. Senthil}
\affiliation{Department of Physics, Massachusetts Institute of Technology, Cambridge MA 02139-4307,  USA}
\date{\today}

\begin{abstract}
Recent experiments have reported chiral time-reversal broken superconductivity in $n$-layer rhombohedral graphene for $n = 4,5, 6$. 
Introducing a moir\'e potential by alignment with a hexagonal boron nitride substrate suppresses the superconductivity but leads instead to various fractional quantum anomalous Hall phenomena. 
Motivated by these observations, we consider the fate of the phase transition between (a chiral) Landau Fermi liquid (LFL) metal and a Composite Fermi Liquid (CFL) metal in the presence of attractive interactions. 
These are parent states, respectively, for the superconductor and the fractional quantum Hall states. 
For weak attractive interactions, the LFL is usually unstable to superconductivity while the CFL is stable. 
This raises the possibility of a direct continuous phase transition between the chiral superconductor and the CFL.
However, we show that generically the LFL close to the transition to the CFL is stable against superconductivity. 
Thus the evolution between the CFL and chiral superconductor goes through an intermediate stable LFL phase for weak attractive interactions. 
With stronger interactions, the evolution can instead go through a non-Abelian paired quantum Hall state.
\end{abstract}

\maketitle

\section{Introduction}
The interplay between topology and strong electron correlations can give rise to a variety of exotic quantum phases, and nowhere is this more apparent than in moir\'e material platforms, in which both of these factors can coexist and compete with one another.
In particular, experiments  examining moir\'e materials such as twisted MoTe$_2$ (tMoTe$_2$) \cite{park2023,cai_nature_2023,xu_prx_2023,anderson_nature_2024} and multilayer rhombohedral graphene \cite{lu2024a,han2024a} aligned with a hexagonal Boron-Nitride (hBN) substrate have found Fractional Quantum  Anomalous Hall (FQAH) states, which exhibit a fractionally quantized quantum Hall conductance without a magnetic field.

Remarkably, experiments on $n$-layer rhombohedral graphene (RnG) for $n = 4,5, 6$ with no moir\'e potential ({\it i.e} no hBN alignment) report \cite{han2025,qin_arxiv_2026} a chiral ({\it i.e} a spontaneously time-reversal broken) superconductor. 
Time-reversal is already broken in the normal metallic state above the superconducting transition. 
Comparing the moir\'eful and moir\'eless devices shows that the parameter range (density and displacement field) in which the superconductor occurs is very similar to that in which the FQAH phenomena occurs, in the presence of the moir\'e potential. 
If we thus imagine gradually dialing up the strength of the moir\'e potential, we will see the chiral superconductor give way to various FQAH states. 

Despite being  two representative paradigms of strongly correlated quantum matter, there is still little known about the interplay of quantum Hall physics and superconductivity. 
Generically, the two phenomena have vastly different origins.
While superconductors are usually described by an instability of a Fermi surface of electrons, quantum Hall states 
require strong interactions and host intrinsic topological order. Recent work \cite{shi_prx_2025,divic2025,shi_scipost_2025, kim_prb_2025,shi_pnas_2025,nosov_prl_2026,pichler2025arxiv,wang2025chiral,han2025a,wang2025}
has revived the old idea of ``anyon superconductivity" and applied it to discuss the possibility of superconductivity in a lightly doped FQAH state. 
However this route is not expected \cite{shi_prx_2025} to be directly applicable to the  chiral superconductor seen in moir\'e-less RnG where a more dramatic change from the FQAH state must occur as the moir\'e potential is turned off \footnote{It may be pertinent to the superconductivity reported in Ref.~\cite{xu2025arxiv} in tMoTe$_2$ close to the 2/3 FQAH state (see  Refs.~\cite{shi_prx_2025,shi_pnas_2025,nosov_prl_2026}).}.

How then does a chiral superconductor evolve into fractional  quantum Hall states? 
As noted above, the chiral superconductor descends from a parent chiral normal metal, which we presume is a  Landau Fermi Liquid (LFL).
In contrast, fractional quantum Hall states descend from a different parent state, namely a Composite Fermi Liquid (CFL) \cite{halperin1993}.
This is a non-Fermi liquid metal with low energy physics determined by a Fermi surface of composite fermions (rather than electrons) coupled to emergent $U(1)$ gauge fields.
Previous work \cite{barkeshli_prb_2012,song_prb_2024} has studied the phase transition between the LFL and CFL states; the transition could be continuous and is predicted \cite{song_prb_2024} to have an interesting universal jump in its resistivity tensor. 

Here we consider the effect of weak attractive interactions on this CFL-LFL transition. 
Naively, the LFL might be expected to be unstable to superconductivity. 
In contrast, due to the gauge interactions, the CFL is stable to composite fermion pairing for weak attractive interactions.
Thus, it is possible that a weak attractive interaction converts the CFL-LFL transition into a CFL-superconductor transition. 
Understanding this transition can then anchor a discussion of the evolution between the superconductor and Jain fractional quantum Hall states that descend from the CFL. 
We will not discuss the microscopic origin \footnote{It is possible that it originates from what microscopically is a repulsive interaction through the Kohn-Luttinger mechanism, or through some other exotic route \cite{kim_prb_2025}.} of the attractive interaction in this paper \cite{guerci2024arxiv,chou_prb_2025,qin_prl_2025,jahin_prb_2026,xu_prl_2025,guerci_prl_2025,chen_nat_2026}.
Instead, we
simply assume that near the Fermi surface, we have an attractive interaction and study its consequences as the metal undergoes the CFL-LFL transition. 

The CFL is one of the best characterized  non-Fermi liquid metals in two-dimensional systems. It does not have an electronic Fermi surface but rather has a Fermi surface of composite fermions, which have been seen in many experimental probes~\cite{shayegan2020,eisenstein1992, willett1993, kang1993, willett1993a, goldman1994, smet1996}. 
While CFLs in conventional quantum Hall systems are stabilized by large magnetic fields, they can also be stabilized by the interplay between Berry curvature and strong interactions in lattice systems in zero field. 
Numerical work on models of tMoTe$_2$ predict a CFL appearing at half-filling of the first (hole) band~\cite{dong2023a, goldman2023}.
Moreover, experimental signatures of the CFL have been observed in tMoTe$_2$~\cite{park2023,anderson_nature_2024}
and in rhombohedral multilayer graphene~\cite{lu_nature_2025, han2025}.

We use the results and framework developed by earlier studies of pairing in non-Fermi liquids \cite{metlitski2015a}
to first analyze the phase transition between the CFL and an ordinary Fermi liquid in the presence of Cooper pairing, with the goal of understanding the behavior of superconducting instabilities near the transition and out of each respective phase.
We find via the renormalization group analysis of~\cite{metlitski2015a} that pairing at the CFL-LFL transition is \textit{suppressed}, implying that the non-Fermi liquid at the CFL-LFL critical point is stable to pairing. 
This suppressed pairing persists into the CFL and LFL phases, leading in particular to a {\it stable} Fermi liquid in the presence of a bare attractive interaction and presenting an obstruction to a direct CFL-superconductor transition. 
Thus for weak attractive interaction, generically a Fermi liquid phase appears between the CFL and the chiral superconductor. 

\begin{figure}[h]
\centering
{\includegraphics[width=0.8\columnwidth]{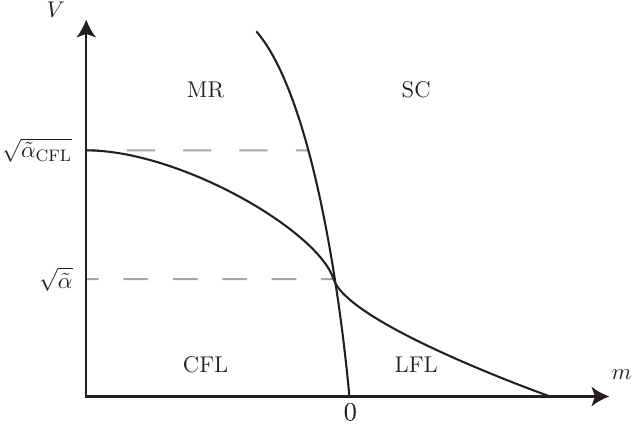}}
\captionsetup{justification=raggedright}
\caption{Phase diagram of the CFL-LFL transition perturbed by $V$, an attractive interaction.
The horizontal axis $m$
represents the tuning parameter across the CFL-LFL transition (equivalently, tuning the mass in the critical theory of Eq.~\eqref{eq:criticaltheory}).
For the idealized case of a Landau level in a periodic potential $U_p$, increasing $m$ corresponds to increasing $U_p$.
The phase diagram assumes that pairing at the critical point slightly favors the LFL phase and $\tilde{\alpha}<\tilde{\alpha}_{CFL}$ (as defined in Eq.~\eqref{eq:dimensionless_alpha}), which are nonuniversal features.
}
\label{fig:phase_diagram}
\end{figure}

Beyond a critical strength of the attractive interaction, the CFL will undergo a pairing transition that produces a non-Abelian fractional quantum Hall state, the Moore–Read state~\cite{moore1991, read2000} (believed to also be realized in Landau level quantum Hall systems at $\nu = 5/2$~\cite{willett1987, radu2008, dolev2008, storni2010}). 
This paired quantum Hall state can have a direct transition into the chiral superconductor. This leads to an alternate pathway between the CFL and the chiral superconductor where the paired quantum Hall state appears as an intermediate phase.
We illustrate a schematic phase diagram in Fig.~\ref{fig:phase_diagram}.
Thus we show that  en route to the superconductor, the CFL can pass through a multicritical point or one of two intermediate phases: either a Moore-Read state with non-Abelian topological order
or a Fermi liquid that is unexpectedly stable to superconductivity due to its proximity to a metallic QCP. Furthermore, we  characterize the universal critical properties of all the associated transitions, including, in particular, the LFL-SC transition which occurs at {\it finite} interaction strength in our system. 

The organization of the paper is as follows. In Section~\ref{sec:criticalTheory}, we review the theory of a phase transition between a composite Fermi liquid and a regular Fermi liquid, as developed in~\cite{barkeshli_prb_2014, song_prb_2024}. 
In Section~\ref{sec:pairing}, we introduce an attractive interaction to the theory and employ the renormalization group to study pairing instabilities of the critical point as well as within the two phases.
We conclude our discussion in Section~\ref{sec:conclusion}.

\section{The CFL-LFL transition}
\label{sec:criticalTheory}
\subsection{Parton Decomposition}
To begin, we will describe the theory of the continuous CFL to LFL phase transition, first discussed in \cite{barkeshli_prb_2014} and later refined in \cite{song_prb_2024}.
Note that the transition considered here requires time reversal symmetry breaking, either explicitly or spontaneous.
In the moir\'e experiments which we are considering, such symmetry breaking often occurs spontaneously as the valley degree of freedom is polarized (in addition to the spin, if it is an independent degree of freedom).
A parton decomposition of the physical electron, $c$, provides a useful framework with which to understand the transition.
We will decompose
\begin{equation}
c=\Phi f,
\end{equation}
where the bosonic parton $\Phi$ and fermionic parton $f$ are both coupled with opposite charge to an emergent $U(1)$ gauge field $a_{\mu}$ introduced by the parton decomposition.
The gauge redundancy fixes the filling of each particle $\nu_c=\nu_{\Phi}=\nu_f$.
Let us consider $\nu_c=1/2$.
We will take $\Phi$ to carry the physical $U(1)$ electric charge. As we are considering this transition occurring in a fully valley polarized state, all the fermions will be effectively spinless.
We are interested in a mean field state in which $f$ occupies the same Landau Fermi liquid state as $c$ in the absence of gauge fluctuations. With this, the low energy theory of the physical electron can be written as 
\begin{equation}
    \label{eq:l_total}\mathcal{L}_{\text{total}}=\mathcal{L}_{FS}[f,-a]+\mathcal{L}[\Phi,a+A]+\cdots,
\end{equation}
where 
\begin{equation}
    \mathcal{L}_{FS}[f,-a]=f^{\dagger}(i\del_t+ia_0+\mu)f+\frac{1}{2m}f^{\dagger}(\nabla +i\vec{a})^2f
\end{equation}
describes a Fermi surface  of $f$ at filling $\nu=1/2$ per unit cell.
Here, $a$ labels the emergent gauge field.
For the boson $\Phi$, we schematically wrote $\mathcal{L}[\Phi,a+A]$
to indicate $\Phi$ couples to $a+A$, where $A$ is a probe gauge field for the physical electric charge.
Within this framework, when $\Phi$ undergoes a bosonic $\nu=1/2$ Laughlin to superfluid transition, the physical electrons undergo a transition from a CFL to LFL.
The last term ``$\cdots$'' in Eq.~\eqref{eq:l_total} includes couplings between the parton sectors, but we will ignore these terms as it is argued \cite{barkeshli_prb_2014,song_prb_2024} that such couplings are likely to be irrelevant.
Consequently, $\Phi$ and $c$ dynamically decouple at the transition and we will simply focus on the critical behavior of $\Phi$, which drives the transition.
We will comment more on this phenomenon later on. 

We will briefly describe the two phases in the parton picture. 
Condensing the boson $\langle\Phi\rangle\ne 0$ places it into a superfluid phase, fixing $a$ to $-A$ at long distances. 
Consequently, the Lagrangian for the Fermi liquid is recovered \footnote{
This is particularly clear in the parton picture, as $\langle \Phi\rangle\ne0$ allows us to identify $f$ and $c$, as the dynamical gauge field $a$ is Higgsed.
The resulting LFL quasiparticle residue will be given by $Z\sim |\langle \Phi\rangle|^2$.
One can also understand this phase by going to the dual vortex picture,
\begin{equation}
    \mathcal{L}_{total}=\mathcal{L}_{FS}[f,-a]+\frac{i}{2\pi}\tilde{a}\wedge d(a+A)+\mathcal{L}[v_{\Phi},\tilde{a}]+\cdots,
\end{equation}
where $\Phi$ carries flux under $\tilde{a}$.
When the boson vortex $v_{\Phi}$ is gapped, integrating out $\tilde{a}$ leads to a mass term for $a_{\mu}$. The resulting Fermi surface state of $f$ is then exactly the ordinary metallic state of the physical electron $c$.}.
Because time reversal symmetry is broken, the LFL state is expected to have orbital loop currents, in addition to a non-zero Hall conductance.

In the case the $\Phi$ is in the $\nu=1/2$ Laughlin state, we can write \cite{barkeshli_prb_2014}
\begin{equation}
    \mathcal{L}_{\Phi}=\frac{1}{2\pi}\tilde{a}\wedge d\tilde{a}+\frac{1}{2\pi}(a+A)\wedge d\tilde{a},
\end{equation}
which is a $U(1)_2$ Chern-Simons theory with an additional gauge field giving the boson current $j_{\Phi}=\frac{1}{2\pi}d\tilde{a}$.
Relabeling $a\rightarrow -a-A$ and integrating out $\tilde{a}$ yields exactly the CFL Lagrangian,
\begin{equation}
    \mathcal{L}_{total}=f^{\dagger}(i\del_t-ia_0-iA_0+\mu)f+\frac{1}{2m}f^{\dagger}(\nabla -i\vec{a}-i\vec{A})^2f+\frac{1}{2}\frac{1}{4\pi}a\wedge da+\cdots.
\end{equation}
Importantly, the resulting state is compressible \cite{barkeshli_prb_2014}, despite the fact that $\Phi$ is in an incompressible Laughlin state. 

There are multiple Lagrangian descriptions of the phase transition of $\Phi$ from the Laughlin state to superfluid, all related by dualities \cite{song_scipost_2023}, but here we will adopt the picture used in \cite{song_prb_2024}. 
We first further fractionalize the boson,
\begin{equation}
    \Phi=d_1d_2,
\end{equation}
into two fermionic partons $d_{1,2}$, in which case the transition from LFL to CFL is described exactly by the fermionic partons undergoing a phase transition from total Chern number $C=0$ to $C=2$.
We assign the charge of the boson probed by $a+A\equiv A_b$ to $d_1$, so $d_2$ is neutral under $A_b$.
In this parton representation both $d_{1,2}$ are at filling $1/2$, glued together by an emergent $SU(2)$ gauge field.
We take a mean field ansatz in which $d_{1,2}$ each see $\pi$ flux through each plaquette, which
doubles the unit cell for the band structure of both $d_{1,2}$ and allows them to form band insulators at half-filling. 
We also take a mean field ansatz that breaks the $SU(2)$ gauge symmetry to a $U(1)$ gauge symmetry, with emergent gauge field $b$.

Let us put $d_{1}$ into a band with Chern number $C_{1}=1$, and consider a band-touching transition changing the Chern number of $d_2$ from $C_2=-1$ to $C_2=1$.
We can describe this transition with two massless Dirac fermions $\psi$ coupled to the gauge field $b$, with background Chern-Simons terms from $C_1=1$ and $C_2=-1$,
 \begin{equation}
    \label{eq:criticaltheory}
\mathcal{L}_b=\sum_{i=1,2}\ol{\psi}_i\slashed{D}_{b}\psi_i
    -m\ol{\psi}\psi
    +\frac{1}{4\pi}(b+A_b)\wedge d(b+A_b)-\frac{1}{4\pi}b\wedge b.
\end{equation}
The above is Pauli-Villar regularized, such that a single massless Dirac fermion has vanishing Hall conductivity 
for $m>0$.
As shown in \cite{barkeshli_prb_2014}, the
superfluid to Laughlin state transition for the boson $\Phi$ occurs when $C_2$ goes from $-1$ to $1$ or equivalently, 
when $m$ goes from positive to negative.
Importantly, Ref.~\cite{song_prb_2024} illustrated that no fine-tuning is needed to ensure that the change in Chern number for $d_2$ is $2$ and not $1$, as would usually be expected in band theory.
Instead, Lieb-Schultz-Mattis \cite{lieb_ann_1961,oshikawa_prl_2000,hastings_2004_prb} constraints that lead to a projective action of translation symmetry on $d_2$ ensure that the flavor adjoint masses, $\ol{\psi}\sigma^a\psi$, are symmetry forbidden \footnote{As argued in \cite{song_prb_2024}, adding a flavor mass to the critical point leads to a trivial insulator.
However, no such insulator of bosons is possible at $\nu=1/2$ unless translation symmetry is broken, leading us to conclude the flavor adjoint masses must break translation}.

\subsection{Critical Field Theory}
We now focus our attention towards the critical point, when $m=0$ in Eq.~\eqref{eq:criticaltheory}.
From inspection, we can see that the critical theory is scale invariant, and we expect it will be described
by a conformal field theory (CFT).
It was argued in \cite{barkeshli_prb_2014} that the transition is continuous.
Specifically, one can show that at the critical point, the bosonic and fermionic parton sectors are essentially decoupled and the effective critical theory is that of a Fermi surface of $f$ coupled to gauge fluctuations.
However, the bosonic sector will affect the low energy gauge fluctuations.
Consequently, its effects will be reflected in corrections to the gauge field propagator.
For completeness, we outline this construction in Appendix~\ref{app:decouple}. 

The field theory at the critical point is a Fermi liquid coupled to a gapless boson $\phi\equiv a_{\perp}$, the component of the emergent gauge field $a$ transverse to the Fermi surface.
This field theoretic framework \cite{mross2010,metlitski2015a} is versatile enough to describe a variety of non-Fermi liquid phases and critical points.
We will allow the fermion $f$ to have a flavor index $\alpha=1,\cdots, N$ (physically in the moir\'e systems under consideration, $N=1$).
The Landau-damped fluctuations of $\phi(\omega,\vec{q})$ interact most strongly with $f_{\alpha}$
in the regions of the Fermi surface (FS) to which $\vec{q}$ is tangent.
We therefore divide the FS into pairs of antipodal patches, labeled
by an index $j$ with directions transverse and tangent to the FS as $x$ and $y$, respectively.
Each patch has width $\Lambda_y\ll k_F$ and thickness $\Lambda_{x}\sim\Lambda_y^2k_F^{-1}\ll\Lambda_y$. We divide $f_{\alpha}$ into patch fields,
\begin{equation}
    f_{\alpha}(\tau,\vec{r})=\sum_{j,s=\pm} f^j_{s\alpha}(\tau,\vec{r})e^{is\vec{k}_j\cdot\vec{r}},
\end{equation}
where we have chosen a representative
FS momentum $\vec{k}_j$ for each patch pair $j$.
We have also assumed in the above an inversion symmetric FS, so each patch pair is located at momenta $s\vec{k}_j=\pm \vec{k}_j$.
While this assumption does not hold in the microscopic model, we do not expect it to affect our conclusions. 
The bosonic patch fields $\phi_j$ are defined to have momentum modes nearly tangent to the FS, with  $q_y<\Lambda_y$ and $q_x<q_y\Lambda_yk_F^{-1}$.
The total action is then a sum of the actions for each FS patch,
\begin{align}
    \label{eq:total_patch_theory}
    S_{total}=\sum_j S_j&=\sum_j S_j[f]+S_j[\phi]+S_j[\phi,f],\\
    S_j[f]&=\sum_{\alpha,s=\pm}\int_{\omega,\vec{q}} f^{j,\dagger}_{s\alpha}(\omega,\vec{q})G_{s}^{-1}
    (\omega,\vec{q})f^{j}_{s\alpha}(\omega,\vec{q}),\quad G_s^{-1}(\omega,\vec{q})=-i\omega+v_F\left(sq_x+\frac{q_y^2}{2K}\right)
    \\
    S_j[\phi]&=\frac{N}{2g^2}\int_{\omega,\vec{q}} |q_y|^{1+\epsilon}|\phi_j(\omega,\vec{q})|^2,
    \\
    \label{eq:gauge_fs_coupling}
    S_j[\phi,f]&=\sum_{\alpha,s=\pm}\int_{\omega,\vec{q},\omega',\vec{q}'} s\lambda \phi_j(\omega',\vec{q}')f^{j,\dagger}_{s\alpha}(\omega+\omega',\vec{q}+\vec{q}')f^{j}_{s\alpha}(\omega,\vec{q}),\quad \lambda=v_F
\end{align}
We have introduced an expansion parameter $\epsilon\sim z_{\phi}-2$, where $z_{\phi}$ is the dynamical critical exponent for the bosonic sector.
Note the bosonic sector also enters through the value of the coupling $g^2$.
This theory can be studied perturbatively by combining a $1/N$ expansion with an $\epsilon$ expansion~\cite{mross2010, metlitski2015a}, although higher loop renormalizations of the gauge coupling $\alpha$
have called into question the validity of this expansion~\cite{ye2022a}. 
In our case of single-component fermions coupled to a gauge field, we have $\epsilon=0$, $N=1$. We also defined the FS curvature $K$, which controls the dispersion of the fermions as the transverse momentum varies. 
In general, $g$, $K$, and $v_F$ will vary along the FS.
Lastly, we observe that $\phi$ couples to the fermion current.
As antipodal patches have opposite Fermi velocities, $\phi$ will couple with opposite sign to antipodal patches, reflected in Eq.~\eqref{eq:gauge_fs_coupling}.

Defining the dimensionless coupling constant
\begin{equation}
\label{eq:dimensionless_alpha}
   \alpha=\frac{g^2\lambda\Lambda_y^{-\epsilon}}{(2\pi)^2}, 
\end{equation}
the renormalization group (RG) flow to one loop order in a large $N$ expansion leads to \cite{mross2010}
\begin{equation}
\label{eq:alpha_rg}
    \beta_{\alpha}=\frac{d\alpha}{dl}=\frac{\epsilon}{2}\alpha-\frac{\alpha^2}{N},\quad \beta_{\lambda}=\frac{d\lambda}{dl}=-\frac{\alpha}{N}\lambda.
\end{equation}
At the fixed point $\epsilon=0$, $\alpha=0$, the system is a marginal Fermi-liquid, with fermion self-energy of the form
\begin{equation}
    \Sigma_f\sim i\omega\log\frac{\Lambda_{\omega}}{\omega}.
\end{equation}
We remark that the CFL phase in the presence of Coulomb interactions is also described by $\epsilon=0$ and $N=1$, though with a different coupling $\alpha_{CFL}$ originating from the Chern-Simons term instead of a universal coupling derived from the CFT critical point.

Currently, we do not have an estimate or physical argument for the expected relationship between the $\alpha$ for the CFL and LFL in a typical experimental system, as $\alpha$ will generically depend very sensitively on microscopic parameters.
Whether one is generically larger than the other based on the CFT origin of the couplings is an open question both from the CFT side and from the microscopics side, as $\alpha$ depends on $g$, coming from the critical boson, and $v_F$.
In the case $\tilde \alpha>\tilde \alpha_{CFL}$, the only change to Fig.~\ref{fig:phase_diagram} is that pairing will be more favored deep into the CFL when compared to the critical point. 
In both cases, the CFL will still be stable to a finite attractive interaction.

\section{Pairing}
\label{sec:pairing}
We assume the presence of some bare attractive electron-electronic interactions, motivated by the empirical presence of a superconducting instability in multilayer rhombohedral graphene systems \cite{han2025}.
Before analyzing the consequences of an attractive interaction, it is useful to first isolate the role of the band structure in the CFL-LFL phase transition.
We adopt an idealized picture by treating the half-filled, nearly flat $C=1$ Chern band of $c$ probed in experiments as a Landau level subjected to a periodic potential $U_p$ with approximately $2\pi$ flux per plaquette. 
This caricature does not apply perfectly to moir\'e systems, which have nonuniform Berry curvature, longer range hoppings, and strong Landau level mixing, but it does allow us to concisely describe a bandwidth tuned transition between the CFL and LFL phases.
However, in some systems, such as twisted MoTe$_2$, the lattice potential tuning the Landau level bandwidth can roughly be connected to the moir\'e potential through a mapping of the continuum model to a Landau level problem \cite{duran_prl_2024,paul_sci_2023,reddy_prb_2023}.
In the presence of real material complications, we still expect our key conclusions, which address universal behavior near the critical point, to still hold as they only depend on the long-wavelength properties of the CFL and LFL near criticality.

We first comment on the fate of the Landau level of $c$ under a periodic potential in the absence of attractive interactions.
In the flat band limit, the CFL phase is realized as $c$ half fills a flat $C=1$ Chern band (or equivalently, $\Phi$ is in a Laughlin state while $f$ realizes a LFL). 
Turning on a periodic potential of $U_p$ with $2\pi$ flux per plaquette, the Landau levels acquire a finite bandwidth.
In the limit that the bandwidth is
large compared to the interaction strength, the kinetic energy is no longer quenched and the resulting state
of $c$ is in the LFL phase.
From another perspective, when the strength of the periodic potential $U_p$ is on the order of the boson gap, $\Phi$ can enter a superfluid state by condensing at the bottom of the band, leading to the LFL phase.

We now wish to understand how the CFL-LFL transition is modified in the presence of a weak attractive interaction. As has been demonstrated in~\cite{metlitski2015a}, critical Fermi surfaces can either enhance or suppress pairing depending on the nature of the interactions.

Despite time reversal symmetry breaking from the Chern band, we will assume a nearly $\vec{k}\rightarrow-\vec{k}$ degeneracy, meaning the pairing instability and Fermi-surface inversion are only weakly perturbed.
We do not expect the details of the Fermi surface anisotropy to affect the universal properties we are after.
In this case, upon adding an attractive interaction, our action remains the same as in Eq.~\eqref{eq:total_patch_theory}, except now with an additional inter-patch perturbation 
\begin{align}
   \delta S\sim\prod_{i=1}^4\int_{\omega_i,\vec{q_i}}f_{\alpha}^{\dagger}(\omega_1,\vec{q}_1)f_{\beta}^{\dagger}(\omega_2,\vec{q}_2)f_{\gamma}(\omega_3,\vec{q}_3)f_{\delta}(\omega_4,\vec{q}_4)(2\pi)^3\delta^3(q_1+q_2-q_3-q_4).
\end{align}
While the above perturbation is between the fermionic partons and not the physical fermions, in the LFL phase, the condensation of $\Phi$ identifies the two. 
In the CFL phase, we expect such a parton-parton interaction to descend from a similar interaction of physical electrons, as at long wavelengths, one can integrate out the gapped $\Phi$ field in the CFL phase.
The resulting gauge invariant electron density is given by $\rho_c=\rho_f$ up to short-wavelength gauge field corrections.
Only the forward scattering
and BCS channels survive the process of renormalization group for the ordinary Fermi liquid, in which the shell of allowed states around the Fermi surface is shrunk under RG. As we 
want to focus on pairing, we will isolate the BCS channel scattering, which means $q_{2,4}=-q_{1,3}$.
Then, we have
\begin{equation}
    \delta S\sim-\frac{1}{4}\prod_{i=1}^2\int_{\omega_i,\vec{q_i}}V^{\alpha\beta\gamma\delta}(\theta_1,\theta_2)f_{\alpha}^{\dagger}(\omega_1,\vec{q}_1)f_{\beta}^{\dagger}(-\omega_1,-\vec{q}_1)f_{\gamma}(\omega_2,\vec{q}_2)f_{\delta}(-\omega_2,-\vec{q}_2)
\end{equation}
where we have taken the interaction to depend only on $\theta_{1,2}$, which are the angles of $q_{1,2}$ that lie on the Fermi surface. 
Defining the dimensionless coupling $\ol{V}=\frac{k_F}{2\pi v_F}V$, the RG flow of the BCS coupling to leading order in $1/N$ is \cite{metlitski2015a} 
\begin{equation}
\label{eq:pairing_rg}
    \beta_{\ol{V}}=\frac{\alpha}{N}-\ol{V}^2.
\end{equation}
Note the above flow is independent of
and closed within the angular momentum channel of $V$.
Therefore, the universal results we derive, including the stability to pairing near criticality, are expected to hold for all angular momentum channels.
The specific channel realized in any system will generically depend on microscopic details.

\subsection{Pairing in the CFL and LFL phases away from criticality}
For an ordinary LFL, $\alpha=0$, and if $\ol{V}<0$ is initially attractive, Eq.~\eqref{eq:pairing_rg} shows there is a runaway flow of $\ol{V}$ to $-\infty$ at a finite $l=l_p$ with the expected BCS instability at energy scale $\Delta_{BCS}\sim \Lambda_{\omega}e^{-1/|\ol{V}|}$.
Furthermore, in the LFL phase, the parton $\Phi$ is condensed, which Higgses the emergent gauge field, and the resulting superconductor is an ordinary electron superconductor, albeit with
an orbital magnetization arising from broken time reversal symmetry.
We expect the exact pairing channel to be determined by which angular momentum channel of $\ol{V}$ diverges first, which is generally a nonuniversal property.
For experimentally relevant moir\'e systems with time reversal symmetry breaking and $C_3$ rotation symmetry in a valley polarized state, the minimal allowed angular momentum pairing is $p\pm ip$.
More exotic cases, such as finite momentum pairing (FFLO) can allow $s$-wave pairing, but we will not consider these in our analysis.
We remark that in the case of chiral $p$-wave pairing, the resulting paired LFL is the standard chiral topological superconductor.
In the case of a highly anisotropic Fermi surface or finite momentum pairing,
the superconducting order may lead to a Bogoliubov Fermi surface, which provides additional gapless sector.

\begin{figure}[h]
\centering
\includegraphics[width=0.7\columnwidth]{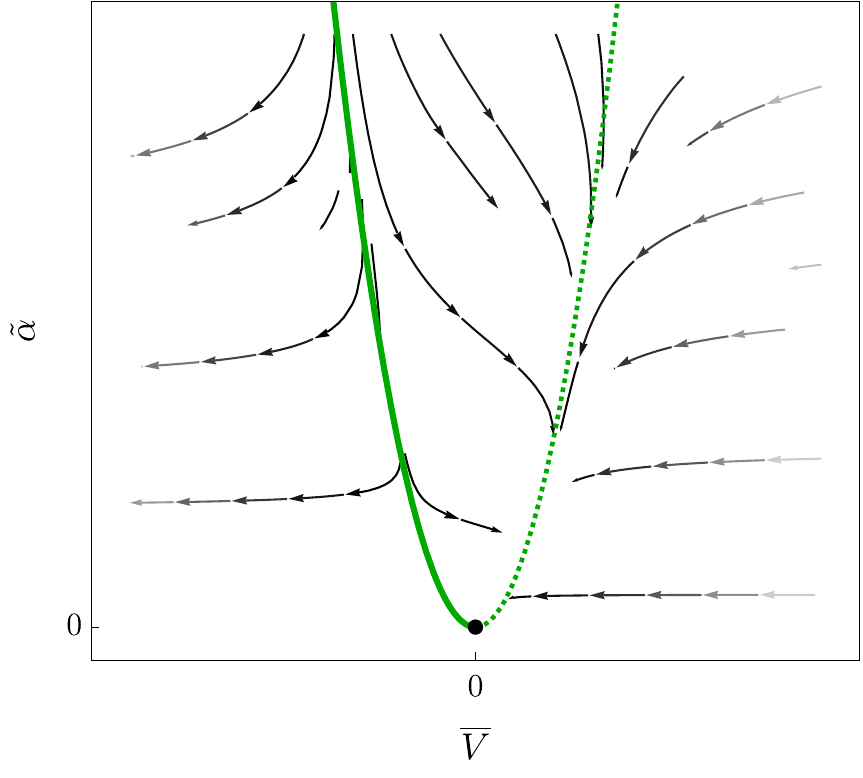}
\captionsetup{justification=raggedright}
\caption{RG flow of the coupling $\tilde{\alpha}$ and attractive interaction $\ol{V}$ for the CFL and critical point.
The solid green curve denotes the separatrix $\ol{V}\sim-\sqrt{\tilde{\alpha}}$, while the dashed green curve shows the attractor line $\ol{V}\sim\sqrt{\tilde{\alpha}}$.
We mention that the separatrix also flows (logarithmically) towards the fixed point $(0,0)$.
}
\label{fig:flow}
\end{figure}

In the CFL phase, we are at the marginal point $\epsilon=0$, at which the coupling $\alpha_{CFL}$ flows logarithmically to $0$ from Eq.~\eqref{eq:alpha_rg}.
As found in \cite{metlitski2015a} and shown in Fig.~\ref{fig:flow}, there is an attractor line at $\ol{V}=\sqrt{\tilde{\alpha}_{CFL}}$ and separatrix $\ol{V}=-\sqrt{\tilde{\alpha}_{CFL}}$ in addition to the fixed point at $\ol{V}=0$, $\tilde{\alpha}_{CFL}=0$,
where we have defined
\begin{equation}
    \tilde{\alpha}_{CFL}\equiv\frac{\alpha_{CFL}}{N}.
\end{equation}
All initial $\ol{V}>-\sqrt{\tilde{\alpha}_{CFL}}$ will flow to the attractor line and then to the fixed point at $\ol{V}=0$, $\alpha=0$.
For initial $\ol{V}<-\sqrt{\tilde{\alpha}_{CFL}}$, there is an instablity and pairing occurs at a finite $l=l_p$ (details are outlined in Appendix~\ref{app:rg_flow}).
Therefore, the CFL is stable to pairing even in the presence of a finite attractive interaction. 
When $\ol{V}<-\sqrt{\tilde{\alpha}_{CFL}}$, pairing will occur, leading to a gap for the fermionic parton $f$.
Note that for pairing close to the critical separatrix, as $\delta\ol{V}=-(\sqrt{\tilde{\alpha}_{CFL}}+\ol{V})\rightarrow 0^+$, the pairing gap exhibits an unusual scaling
\begin{equation}
    \Delta_{pair,CFL}\sim \Lambda_{\omega}e^{-l_p}\sim  \Lambda_{\omega}\delta\ol{V}^{-\frac{1}{16}\log\delta\ol{V}}.
\end{equation}

Let us take $p+ip$ pairing for concreteness, which is always a possibility in the CFL phase as the Chern-Simons term mediates a charge-current interaction that leads to a $p+ip$ pairing instability \cite{greiter_prl_1991,greiter_nucl_1992}.
On the CFL side, pairing of the composite fermions gaps the neutral fermions, while the charged sector of $\Phi$ remains in the $U(1)_2$ Laughlin state.
Gauging the parton $\mathbb{Z}_2$ leads to
the Moore-Read state, with $[\mathrm{Ising}\times U(1)_8]/\mathbb{Z}_2$ topological order and anyons labeled by $(x,j)$, where $x\in\{1,\psi,\sigma\}$, $j\in\mathbb{Z}_8$, and $(\psi,j)\sim(1,j+4)$.
Furthermore, $j$ is even if $x = (1,\psi)$, and $j$ is odd if $x = \sigma$.
The charge $e/2$ semion ($j=2$) in the Moore-Read state is precisely the descendant of the charge $e/2$ semion of the Laughlin state of $\Phi$.

Tuning $\Phi$ across the Laughlin to superfluid transition drives a transition from the Moore-Read phase to an electronic $p+ip$ superconductor.
To see this, note that from the perspective of the Moore-Read state, only the charge sector becomes critical; the transition of $\Phi$ closes the charge gap while the neutral Ising sector remains gapped.
Equivalently, the semion of Moore-Read state goes gapless, Higgsing the internal Chern-Simons gauge field and eliminating the 
Abelian $U(1)_8$ topological order of the charged sector.
The transition leaves intact the gapped neutral Ising sector (arising from $p+ip$ pairing of the composite fermions).
One $\Phi$ is in a superfluid phase,
we glue the neutral Ising sector to $\Phi$ and obtain a $p+ip$ superconductor of physical electrons, as claimed.
Note that it is not enough to simply condense the unit charge boson within the Moore-Read phase, as doing so confines the Ising sector and leads to a trivial superconductor of electrons. Taking a different $l=n$ pairing channel (such as in some cases where $f$ wave pairing may occur \cite{chou_prb_2025,qin_prl_2025}) yields a similar story, except that the pairing of the composite fermions in the CFL phase yields a generalized Moore-Read state, with a modified chiral central charge $c_-=1+n/2$.

\subsection{Pairing at the CFL-LFL critical point}

\begin{figure}[h]
\centering
\includegraphics[width=0.7\columnwidth]{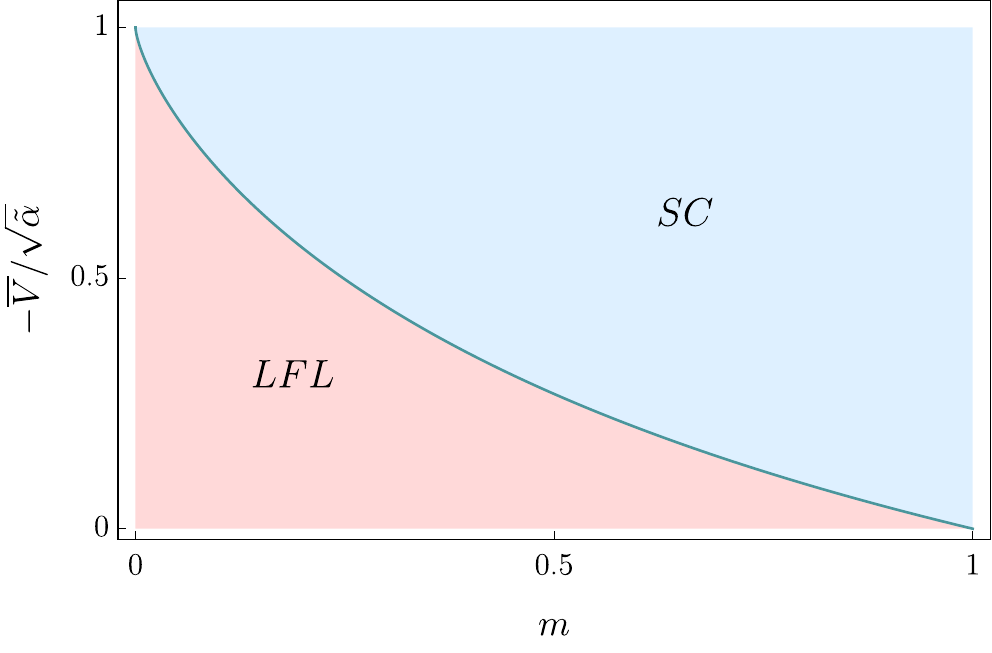}
\captionsetup{justification=raggedright}
\caption{Stability of the LFL state to pairing close to the phase transition transition in terms of the bare attractive interaction $\ol{V}$, from Eq.~\eqref{eq:fl_bcs_stability}.
The horizontal axis is normalized in units of $\Lambda_{\omega}$.
}
\label{fig:fl_bcs_boundary}
\end{figure}

Now let us examine the regimes in the vicinity of the critical point. 
At the critical point with coupling $\tilde{\alpha}$, the RG flow is the same as in the CFL (Fig.~\ref{fig:flow}).
We observe
that even when $0>\ol{V}>-\sqrt{\tilde{\alpha}}$, the LFL will still be stable to pairing if the system is close enough to the critical point.
To see this, we argue as in \cite{metlitski2015a}.
Imagine tuning the fermion mass in Eq.~\eqref{eq:criticaltheory} to $m>0$ so that $\Phi$ is in the superfluid phase.
The gauge field $a$ then becomes Higgsed by the superfluid condensate below a momentum scale
$q_*\sim |m|^{\nu}$.
The correlation length exponent $\nu$ is estimated in Eq.~\eqref{eq:corr_length_exp} to be $\sim 1.42$ by a large $N_f$ calculation (here $N_f=2$ characterizes the physical bosonic CFT).
Then, below the energy scale $\epsilon_*\sim q^2$, gauge fluctuations are not critical, and the system behaves like an ordinary LFL.
Note $\epsilon_*$ is also an IR cutoff for the renormalization group flows for $\ol{V}$ and $\alpha$.
At scales above $\epsilon_*$, $\ol{V}$ and $\tilde{\alpha}$ flow toward the attractor line $\ol{V}=\sqrt{\tilde{\alpha}}>0$.
By the time $\epsilon_*$ is reached in the RG flow, we can have
$\ol{V}(\epsilon_*)>0$, and no pairing will occur as energy is lowered below $\epsilon_*$ and the system enters the ordinary LFL regime.
Therefore, in the vicinity of the critical point, the Fermi liquid will be stable to a finite attractive interaction.
We can make this more precise by integrating  Eq.~\eqref{eq:pairing_rg}, from which we obtain (details in Appendix~\ref{app:rg_flow})
that at a distance $|m|$ away from the critical point into the Fermi liquid phase, the Fermi liquid is stable to a bare pairing attraction of scale
\begin{equation}
   \ol{V}>\tanh\left[\frac{2}{\sqrt{\tilde{\alpha}}}
        -2\sqrt{\frac{1}{\tilde{\alpha}}+\log\left(\frac{\Lambda_{\omega}}{|m|^{2\nu}}\right)}\right]\sqrt{\tilde{\alpha}},
\end{equation}
in the limit of small $\tilde{\alpha}$.
This defines the region of stability of the  LFL to BCS pairing proximate to criticality, shown in Fig.~\ref{fig:fl_bcs_boundary}.
Setting the inequality above to an equality gives us $\ol{V}_*$, the critical bare attraction.
Note that as expected, deep in the LFL phase we have $\ol{V}_*\rightarrow 0$, while near the critical point we recover $\lim_{m\rightarrow 0}\ol{V}_*=-\sqrt{\tilde{\alpha}}$.

In the LFL phase close to criticality, we can also extract the scaling of the gap in terms of $\delta\ol{V}\sim \ol{V}-\ol{V}_*$.
We have defined $\ol{V}_*$ to be the critical bare attraction past which the LFL experiences a BCS instability.
In the LFL phase, we have that
\begin{equation}
    \Delta_{BCS}\sim \Lambda_{\omega} e^{-1/\ol{V}(\epsilon_*)}
\end{equation} where $\ol{V}(\epsilon_*)$ is the result of RG flowing the bare coupling $\ol{V}$ to $\ol{V}(\epsilon_*)$, (assuming that $\ol{V}(\epsilon_*)<0$ so that there is a BCS instability).
We find that $\ol{V}(\epsilon_*)\propto \delta\ol{V}$, so that
\begin{equation}
   \Delta_{BCS}\sim\Lambda_{\omega}e^{-\frac{f(m)}{\delta \ol{V}}}, 
\end{equation}
where the function $f(m)\rightarrow 1$ as $m\rightarrow -\infty$, which recovers the expected result deep in the LFL phase.
Therefore, even near criticality, the LFL BCS gap retains its familiar exponential form as a function of 
$\delta\ol{V}$.

Close to the critical point on the CFL side, when we tune $m>0$, the CFL is also stable to pairing. This is because although below an energy scale $\epsilon_*\sim m^{2\nu}$, there will be no more screening of the gauge field $a$ by the critical bosonic sector, and the system below $\epsilon_*$ is in a CFL state.
The action involves a Chern-Simons term and will have an effective coupling $\tilde{\alpha}_{CFL}$.
As before, at scale $\epsilon_*$, $\ol{V}$ will approach the attractor line $\ol{V}=\sqrt{\tilde{\alpha}}>0$.
Therefore, no pairing instability will occur on the CFL side of the transition as well.
However, the stability of the system to pairing deep in the CFL compared to the critical point will be affected by the relative value of $\tilde{\alpha}$ and $\tilde{\alpha}_{CFL}$.
Because the threshold bare interaction needed to induce pairing at the critical point is $\ol{V}_*=-\sqrt{\tilde{\alpha}}$, we observe that
if $\tilde{\alpha}<\tilde{\alpha}_{CFL}$,
pairing in the pure CFL phase will require a bare attractive attraction of strength $\overline{V}_*\leq -\sqrt{\tilde{\alpha}_{CFL}}<-\sqrt{\tilde{\alpha}}$. Therefore, in the case $\tilde{\alpha}<\tilde{\alpha}_{CFL}$, superconductivity will be enhanced at criticality relative to deep in the CFL phase (and vice versa for $\tilde{\alpha}_{CFL}<\tilde{\alpha}$). 
Note that the relative magnitude of $\tilde{\alpha}$ and $\tilde{\alpha}_{CFL}$ is not a universal 
characteristic but will depend on microscopic details.
We also remark that the stability of the CFL to pairing in the vicinity of the transition still holds if there is no Coulomb interaction but instead only a short range interaction $U(x)\sim|x|^{-2}$ \footnote{
This corresponds to $\epsilon=1$, in which the fixed point at $(\ol{V},\tilde{\alpha})=(0,0)$ splits into fixed points at $\ol{V}=\pm\sqrt{\epsilon/2}$.
The fixed point $\ol{V}=-\sqrt{\epsilon/2}$
is unstable and $\ol{V}=\sqrt{\epsilon/2}$ is a stable fixed point, controlling the CFL phase.
Therefore, with short range interactions, the CFL phase is still stable to pairing.
Furthermore, the CFL in the proximity of the critical point is also stable to weak pairing,
with the same reasoning as before.
Even a distance away from the critical point into the CFL phase, there is an energy scale $\epsilon_*\sim m^{2\nu}$ below which the system is described by the CFL theory.
Above $\epsilon_*$
the flow of $\ol{V}$ tends to the attractor line $\ol{V}=\sqrt{\tilde{\alpha}}$.
Then, if $\ol{V}(\epsilon_*)>-\sqrt{\epsilon/2}$ at the crossover scale, no pairing will occur as we lower energy into the CFL regime.}.

Lastly, we analyze if an attractive interaction at the critical point favors the CFL or LFL phase.
Generically, an infinitesimal BCS coupling will contribute counterterms in the action that may favor the CFL or LFL.
However, we claim that while this effect is present, it is nonuniversal.
To see this, we can focus on the bosonic parton sector, which tunes the transition.
Specifically, how will an 
infinitesimal attractive density-density interaction of $\Phi$ (descended from an interaction term for the electron $c$) renormalize the mass in Eq.~\eqref{eq:criticaltheory}?
Exactly at the (massless) critical point, the attractive interaction will generically contribute four fermion terms, in addition to more irrelevant contributions,
\begin{equation}
\label{eq:v_perturb}
    \ol{V}(\overline{\psi}M\psi)(\overline{\psi}N\psi)+\cdots
\end{equation}
as dictated by symmetry.
In fact, simply from the fact that $\ol{V}$ is time-reversal symmetric, it cannot renormalize the chiral mass $m\ol{\psi}\psi$, and it seems the critical point will be equally stable aginst going into CFL or LFL phase.
It might naively seem that the absence of a chiral mass renormalization is a non-perturbative result that relies only on the symmetry of the critical point.
However, as the critical theory 
(Eq.~\eqref{eq:criticaltheory}) is chiral, irrelevant chiral operators will generically induce a bare mass $\delta_m$ even at the massless fixed point.
More accurately, the massless fixed point (Eq.~\eqref{eq:criticaltheory}) is described by
\begin{equation}
\mathcal{L}_b=\sum_{i=1,2}\ol{\psi}_i\slashed{D}_{b}\psi_i+\delta_m\ol{\psi}\psi+(chiral\;terms),
\end{equation}
where $\delta_m$ is tuned to 
criticality and $(chiral\;terms)$ include irrelevant contributions (e.g., $(\ol{\psi}\psi)(\ol{\psi}M\psi)$, $(\ol{\psi}\psi)f^{\mu\nu}f_{\mu\nu}$, $(\ol{\psi}\psi)^3$, etc.).
An attractive interaction as in Eq.~\eqref{eq:v_perturb} will shift the mass at order $\ol{V}\cdot\delta_m$.
The sign of this renormalization depends on $\delta_m$, which is nonuniversal since it arises from irrelevant time-reversal symmetry breaking couplings at the critical point. 
Consequently, a pure attractive interaction at criticality can drive the system into either the CFL or LFL phase, depending on microscopic (nonuniversal) details.

In conclusion, we obtain the phase diagram in Fig.~\ref{fig:phase_diagram}.
Note that the analysis above can be readily applied in the presence of additional or different tuning parameters such as coupling to nematic order.
We briefly comment on this in Appendix~\ref{app:nem}.

\section{Conclusion}
\label{sec:conclusion}
Recent experiments on rhombohedral multilayer graphene~\cite{han2025,qin_arxiv_2026} find a chiral superconductor in a range of density/displacement field in the moir\'eless limit. 
Turning on a moir\'e potential by aligning with a hBN substrate suppresses the superconductivity but leads instead to fractional quantum anomalous Hall phenomena in the same parameter range. Thus in this system, increasing the strength of the periodic potential leads to an evolution between the chiral superconductor and FQAH states. 
In R4G, the most prominent superconductor in the moir\'eless limit is replaced by the CFL in the moir\'e-full system.  The chiral superconductor itself emerges at low-$T$ out of a chiral metal normal state, presumed to be a Landau Fermi liquid. 

As we discussed earlier, in the idealized case of a Landau level in a periodic potential, tuning across the CFL-LFL phase transition corresponds to increasing $U_p$.
In more general experimental platforms, as far as universal physics is concerned, any experimental knob that tunes through the CFL-LFL phase transition will correspond in the low energy theory to changing the sign of mass.
For example, in rhombohedral graphene aligned with hBN, one can tune a displacement field $D$.
Another possible tuning parameter could be increasing pressure \cite{yankowitz_science_2019,wang_prl_2025}.

Motivated by these observations, we studied the possible pathways from the chiral superconductor to the CFL by considering the effect of an attractive interaction on the previously studied CFL-LFL transition.    
We find that, similar to the behavior within the composite Fermi liquid phase, the gauge fluctuations at the critical point work to suppress pairing, leading to a the CFL-LFL fixed point being stable to weak attractive interactions, as illustrated in Fig.~\ref{fig:phase_diagram}. This stability extends to a finite extent into the Fermi liquid phase, leading to a region in parameter space hosting a metallic system stable against superconductivity. A stronger attractive interaction will lead to pairing in both the CFL and the LFL. The former becomes the non-Abelian Moore-Read quantum Hall state which thus has a continuous phase transition to a chiral superconductor. 

We thus conclude that generically, the path from the CFL to the superconductor will proceed through an intermediate phase which is either a Landau Fermi liquid, or the Moore-Read quantum Hall state. 
Both of these possibilities are intriguing as the former is stable against superconductivity despite being a metallic system, while the former realizes non-Abelian topological order. 
By analyzing the renormalization group flow of the attractive interactions, we have characterized the shape of this stable region, as well as the nature of the zero temperature phase transition between Fermi liquid and superconductor.

While we have found that the CFL-LFL critical theory generically works to suppress pairing, the possibility of engineering gapless degrees of freedom to instead \textit{enhance} pairing and thereby provide a route to non-Abelian topological order is of great interest. 
As discussed in Appendix~\ref{app:nem}, critical nematic fluctuations (though fine-tuned) provide one such path. 
The experimental evidence in rhombohedral multilayer graphene of pairing existing proximate to a CFL also motivates more extensive analysis of the behavior of pairing in non-Fermi liquids through methods other than the specific expansion utilized in~\cite{metlitski2015a}. Complementary numerical studies of lattice models realizing the CFL-LFL transition would provide more quantitative insight into the phase transition and critical theory we have explored here.

\section{Acknowledgments}
We thank Tonghang Han, Long Ju, and Max Metlitski for useful discussions. Y.Z. was supported by the National Science Foundation Graduate Research Fellowship
under Grant No. 2141064. L.S was supported by the MIT Pappalardo Fellowship. T.S. was supported by the Department of Energy under grant DE-SC0008739. 
\appendix

\section{Dynamical Decoupling}
\label{app:decouple}
To obtain an effective field theory of the critical point, let us describe the effect of the couplings between the fermion, bosons, and gauge field at criticality (when the bosonic sector is in a CFT).
In the absence of gauge field fluctuations, a generic interaction between the boson and fermion sectors involves coupling an operator $\mathcal{O}$ from the boson sector to a particle-hole excitation near the Fermi surface.
Integrating out the fermions leads to a Landau damped term,
\begin{equation}
\label{eq:landau_damp}
    \sim\int_{\omega,\vec{q}} \frac{|\omega|}{|\vec{q}|}|\mathcal{O}(\vec{q},\omega)|^2,
\end{equation}
for small $\omega \ll q$.
At the critical point, Eq.~\eqref{eq:criticaltheory}, the most relevant operator is $\mathcal{O}={\Phi}^{\dagger}\Phi$, which has scaling dimension $3-\nu^{-1}$.
In a large $N_f$ expansion (here $N_f=2$) of Eq.~\eqref{eq:criticaltheory}, the correlation length exponent $\nu$ of the bosonic CFT was found to be \cite{chen_prb_1993}
\begin{equation}
\label{eq:corr_length_exp}
\nu^{-1}\sim 0.705+\mathcal{O}\left(\frac{1}{N_f}\right)^2\implies \nu\sim 1.42>\frac{2}{3},
\end{equation}
which suggests that the Landau damped perturbation Eq.~\eqref{eq:landau_damp} is irrelevant.
The other operator that must be irrelevant is a charge density wave in the boson sector, which leads to a term 
\begin{equation}
\label{eq:landau_damp2}
    \sim\int_{\omega,\vec{q}} |\omega|\cdot|\mathcal{O}_{CDW}(\vec{q},\omega)|^2.
\end{equation}
This will be irrelevant at the critical
point if the scaling dimension of $\mathcal{O}_{CDW}$ is greater than $1$, $\Delta_{CDW}>1$. In the
free Dirac fermion theory, $\Delta_{CDW}=2$. Incorporating gauge fluctuations, the scaling dimension
will be reduced, but
we will assume $\Delta_{CDW}>1$ so that Eq.~\eqref{eq:landau_damp2} is irrelevant. 

Therefore, the boson and fermion sectors are essentially decoupled, and the CFL-LFL transition is continuous in the absence of gauge fluctuations.
Gauge fluctuations are not expected to change the continuous nature of the critical point \cite{senthil_prb_2008,barkeshli_prb_2014}.
From boson sector, the gauge fluctuations only lead
to an analytic corrections to the boson propagator
and do not alter the critical singularities structure of the self energy.
The Fermi surface leads to a Landau damped contribution 
\begin{equation}
\delta\mathcal{L}\sim\frac{\omega}{q}a^{\mu}a_{\mu}.
\end{equation}
At the mean field level, the critical bosons have dynamical critical exponent $z=1$, so that Landau damping acts as a Higgs mass and suppresses gauge fluctuations in the boson sector \cite{senthil_prb_2008}.

On the contrary, gauge fluctuations can significantly alter the Fermi surface theory.
Tangentially along the Fermi surface, $q$ and $\omega$ do not scale with the same power.
The momentum normal to the Fermi surface scales $\sim \omega$ while the tangential momenta scales $\sim\omega^{1/2}$, so gauge fluctuations can transfer tangential momentum to the fermions.
We note the decoupling of the boson and fermion sector (and continuous nature of the phase transition) is not destroyed even by the guage field fluctuations.
This is because, as argued in \cite{senthil_prb_2008}, 
the gauge field does not affect the form of the fermion propagator at small $q$, so the form of couplings like Eq.~\eqref{eq:landau_damp} is unchanged.

Therefore, at the critical point theory we have two decoupled systems; the Fermi surface with gauge fluctuations and then the boson CFT.
We can ignore the boson critical theory, which only affects the fermion sector through the gauge field fluctuations.
Integrating out the gapless bosonic sector at the critical point, we obtain the effective action for the gauge field
\begin{equation}
    S_a\sim\int_{\omega,\vec{q}}  q|a_{\perp}(\omega,\vec{q})|^2
\end{equation}
In the above, we have gauged fix with the Coulomb gauge, so that the longitudinal $a_0$ and the transverse spatial component $a_{\perp}=(\hat{z}\times\hat{k})\cdot\vec{a}$ are decoupled.
The coupling of $a_0$ to the fermion density is Debye-screened and can be ignored.
We remark that in the CFL phase with a Coulomb interaction, the same term $q|a_{\perp}|^2$ is generated, albeit with a different proportionality/coupling constant from the critical point.

\section{Solution to the RG equations with attractive interactions}
\label{app:rg_flow}
We will follow and extend the analysis in \cite{metlitski2015a}.
Defining $\ol{V}=v\sqrt{\tilde{\alpha}}$ and using $\beta_{\tilde{\alpha}}=-\tilde{\alpha}^2$, we have the RG flow equation for $v$,
\begin{equation}
    \beta_v=\sqrt{\tilde{\alpha}}(1-v^2)+\frac{\tilde{\alpha}v}{2}.
\end{equation}
In the regime approaching the transition, we will take the limit of $\tilde{\alpha}\rightarrow 0$, so the latter term above is subleading and we will neglect it for our analysis.
Dividing both sides by $\beta_{\tilde{\alpha}}$, we obtain
\begin{equation}
    \dfrac{dv}{d\tilde{\alpha}}=-\tilde{\alpha}^{3/2}(1-v^2).
\end{equation}
Integrating the above equation from the bare values $(\tilde{\alpha}_0,v_0)$, we obtain
\begin{equation}
    \label{eq:vl}
    v(l)=\dfrac{(v_0+1)e^{4(\tilde{\alpha}(l)^{-1/2}-\tilde{\alpha}_0^{-1/2})}+(v_0-1)}{(v_0+1)e^{4(\tilde{\alpha}(l)^{-1/2}-\tilde{\alpha}_0^{-1/2})}-(v_0-1)}.
\end{equation}
As discussed in the main text, if $v_0>-1$, then $v$ flows to $1$ and $\ol{V}$ flows to $\sqrt{\tilde{\alpha}}$.
If $v_0=-1$, then we are at the fixed transition line.
Finally, as if $v_0<-1$, the $\ol{V}$ flows to $-\infty$ and diverges at $l=l_p$.
To find $l_p$, the denominator of Eq.~\eqref{eq:vl} must vanish, which means 
\begin{align}
    (v_0+1)e^{4(\tilde{\alpha}(l_p)^{-1/2}-\tilde{\alpha}_0^{-1/2})}-(v_0-1)=0
    \implies 
    \tilde{\alpha}(l_p)^{-1/2}=\tilde{\alpha}_0^{-1/2}+\frac{1}{4}\log\left(\dfrac{v_0-1}{v_0+1}\right)
\end{align}
Using 
\begin{equation}
    \tilde{\alpha}(l)=\dfrac{\tilde{\alpha}_0}{1+\tilde{\alpha}_0l}
\end{equation}
from $\beta_{\tilde{\alpha}}=-\tilde{\alpha}^2$, we obtain
\begin{equation}
    l_p=\frac{1}{16}\left(\log\left(\dfrac{\ol{V}-\sqrt{\tilde{\alpha}}}{\ol{V}+\sqrt{\tilde{\alpha}}}\right)
    +4
    \tilde{\alpha}^{-1/2}\right)^2-\tilde{\alpha}^{-1},
\end{equation}
entirely in terms of the bare values (we have omitted the $0$ subscript for ease of notation).
Near the separatrix, as $\delta\ol{V}=-(\sqrt{\tilde{\alpha}}+\ol{V})\rightarrow 0^+$, the pairing gap exhibits an unusual scaling form
\begin{equation}
    \Delta\sim \Lambda_{\omega}e^{-l_p}\sim  \Lambda_{\omega}\delta\ol{V}^{-\frac{1}{16}\log\delta\ol{V}}.
\end{equation}
In the LFL phase proximate to the transition, we can analyze the strength of the bare attraction needed to trigger the BCS instability.
As in the main text, suppose we tune to a fermion mass $m>0$.
Then, the question of whether the LFL is stable is equivalent to whether $\ol{V}(\epsilon_*)>0$, where 
$\epsilon_*\sim |m|^{2\nu}$ is the IR cutoff for the RG flow (recall $\nu$ is estimated in Eq.~\eqref{eq:corr_length_exp}). 
Intuitively, $\epsilon_*$ is the energy scale above which the gauge fluctuations are still critical.
To find this threshold $\ol{V}$, we note that for a fixed $l$ (assuming $l_p$ has not been reached yet), the BCS instability will occur when
\begin{equation}
    v_0=-\tanh\left(2\cdot\left(\tilde{\alpha}(l)^{-1/2}-\tilde{\alpha}^{-1/2}\right)\right).
\end{equation}
In terms of the scale $l$, $\ol{V}(l)\ge 0$ when 
\begin{equation}
    l\ge -\dfrac{\operatorname{ArcTanh}(v_0)}{\sqrt{\tilde{\alpha}}}+\dfrac{\operatorname{ArcTanh}(v_0)^2}{4}.
\end{equation}
Physically, the closer $v_0$ is to $-1$, the deeper into the IR one must flow in order to avoid the pairing instability (the closer to the critical point one must be).
In terms of $|m|$, we have $\epsilon_*\sim \Lambda_{\omega}e^{-l_{FL}}\sim |m|^{2\nu}$, so that
\begin{equation}
\label{eq:fl_bcs_stability}
    |m|<\left(\Lambda_{\omega}\exp\left(\dfrac{\operatorname{ArcTanh}(v_0)}{\sqrt{\tilde{\alpha}}}-\dfrac{\operatorname{ArcTanh}(v_0)^2}{4}\right)\right)^{\frac{1}{2\nu}}
\end{equation}
is the condition for the LFL to be stable to pairing.
This is plotted in Fig.~\ref{fig:fl_bcs_boundary}.
An equivalent way of viewing Eq.~\eqref{eq:fl_bcs_stability} is to observe that at a fixed $m$, the Fermi liquid will be stable if the bare attractive interaction satisfies
\begin{equation}
    \ol{V}>-\tanh\left(2\cdot\left(\tilde{\alpha}(\epsilon_*)^{-1/2}-\tilde{\alpha}^{-1/2}\right)\right)\sqrt{\tilde{\alpha}}.
\end{equation}
Near the critical point, this becomes the condition
\begin{equation}
\ol{V}>\tanh\left[\frac{2}{\sqrt{\tilde{\alpha}}}
        -2\sqrt{\frac{1}{\tilde{\alpha}}+\log\left(\frac{\Lambda_{\omega}}{|m|^{2\nu}}\right)}\right]\sqrt{\tilde{\alpha}},
\end{equation}
which yields the behavior $\ol{V}>-\sqrt{\alpha}$ as expected in the limit $m\rightarrow0$.
\section{Pairing in the presence of nematic fluctuations}
\label{app:nem}
Even though experimentally, the presence of nematic fluctuations is not well established,
we briefly comment on how the phase transition is modified in the presence of a coupling to a nematic order parameter, $\varphi$.
In the physical case $N=1$ we are focusing on, $\varphi$ could in actuality be time reversal (and inversion) symmetry breaking as the fermions themselves are valley polarized.
Nematic fluctuations will add the following terms to the action in Eq.~\eqref{eq:total_patch_theory},
\begin{align}
    S_j[\varphi]&=\frac{N}{2g_{\varphi}^2}\int_{\omega,\vec{q}} |q_y|^{1+\delta}|\varphi^j(\omega,\vec{q})|^2,
    \\
    S_j[\varphi,f]&=\sum_{\alpha,s=\pm}\int_{\omega,\vec{q},\omega',\vec{q}'} \lambda_{\varphi} \varphi^j(\omega',\vec{q}')f^{j,\dagger}_{s\alpha}(\omega+\omega',\vec{q}+\vec{q}')f^{j}_{s\alpha}(\omega,\vec{q}).
\end{align}
For ordinary nematic fluctuations of the Hertz-Millis type, $\delta=1$, but like $\epsilon$, $\delta$ also serves as an expansion parameter for the RG procedure.
The above allows for a term $r\varphi^2$; we assume we are at the critical point, so we have tuned $\varphi$ to be massless.
We note that because a $U(1)$ rotation of $\varphi$
corresponds to a spatial rotation, 
$\lambda$ should contain an angular dependence on the angle between $\vec{q}'$ and $\vec{q}$.
However, we will ignore this as such details will not affect
our conclusions about the instabilities to pairing, though it will certainly be important in determining which angular momentum channel pairing will occur in.
Lastly, we remark
that various types of possible orbital orders in moir\'e systems was described in \cite{xu_prb_2020}, in which it is noted that there is an allowed $r(\varphi^3+\varphi^{*^3})$, compatible with $C_3$ symmetry in traditional moir\'e systems.
Notably, such a cubic term contributes to the bosonic self energy a contribution scaling as
\begin{equation}
\delta\Sigma_{\varphi}(\omega,\vec{q})\sim r^2\sqrt{\omega^{2/3}+cq^2}\sim |q|    
\end{equation}
in the low frequency limit, corresponding to $\delta=0$.
This modifies the fermionic self energy to take the marginal Fermi liquid form, 
\begin{equation}
    \Sigma_f(\omega,\vec{q})\sim i\omega\log\frac{\Lambda_{\omega}}{\omega},
\end{equation}
instead of the standard form, $\Sigma_f\sim i\operatorname{sgn}(\omega){\omega}^{2/3}$, that is expected for nematic critical points in the Hertz-Millis framework.
Therefore, lattice symmetries can greatly affect the critical singularities at the nematic critical point.
In the presence of nematic fluctuations, we then have the renormalization group equations
\begin{align}
    \beta_{\alpha}&=\frac{\epsilon}{2}\alpha-\frac{\alpha\cdot\alpha_{+}}{N},
    \quad \beta_{\lambda}=-\frac{\alpha_{+}}{N}\lambda\\
    \beta_{\alpha_{\varphi}}&=\frac{\delta}{2}\alpha_{\varphi}-\frac{\alpha_{\varphi}\cdot\alpha_{+}}{N},
    \quad \beta_{\lambda_{\varphi}}=-\frac{\alpha_{+}}{N}\lambda_{\varphi}\\
    \beta_{\ol{V}}&=\frac{\alpha_{-}}{N}-\ol{V}^2,
\end{align}
as a straightforward generalization of \cite{mross2010}, also considered in \cite{mesaros_prb_2017}.
In the above, we have defined the dimensionless coupling constant $\alpha_{\varphi}$ as before, 
\begin{equation}
    \alpha_{\varphi}=\frac{g_{\varphi}^2\lambda\Lambda_y^{-\delta}}{(2\pi)^2},
\end{equation}
in addition to 
\begin{equation}
\alpha_{\pm}=\alpha\pm\alpha_{\varphi}.
\end{equation}
When  $\epsilon,\delta>0$, there is an unstable fixed point at $(\alpha,\alpha_{\varphi})=(0,0)$, in addition to the fixed points $(\alpha,\alpha_{\varphi})=(\alpha_{*},0)$ and $(0,\alpha_{\varphi*})$ where $\alpha_{*}=N\epsilon/2$ and $\alpha_{\varphi*}=N\delta/2$.
Note the case for which $\epsilon=\delta$ is quite fine tuned, as microscopic details such as filled Landau levels can cause $\epsilon$ and $\delta$ to vary from their bare values. 
However, when $\epsilon=\delta$, there is a line of fixed points  connecting $(\alpha_{*},0)$ to $(0,\alpha_{\varphi*})$.

For the physical case we have specified in which $\epsilon=\delta=0$, there is only a single fixed point at $(\alpha,\alpha_{\varphi})=(0,0)$ that is stable, and we obtain the same behavior of pairing at the critical point, except with an effective interaction $\alpha_-$.
Notably, we observe sufficiently strong nematic fluctuations can drive the interaction to be attractive, similar to the case of the pure Ising-nematic QCP. 
Furthermore, from the similar arguments to the case of pure gauge fluctuations, the superconducting instability will survive even when one tunes slightly away from the critical point.
More details on the interplay between nematic and gauge fluctuations is considered in \cite{mesaros_prb_2017}, and we will not further consider it in the case of the CFL-LFL transition as the resulting critical point is fine-tuned, requiring both the nematic and gauge sectors to simultaneously go critical.

\bibliography{refs,leynaRefs}

\begin{thebibliography}{69}%
\makeatletter
\providecommand \@ifxundefined [1]{%
 \@ifx{#1\undefined}
}%
\providecommand \@ifnum [1]{%
 \ifnum #1\expandafter \@firstoftwo
 \else \expandafter \@secondoftwo
 \fi
}%
\providecommand \@ifx [1]{%
 \ifx #1\expandafter \@firstoftwo
 \else \expandafter \@secondoftwo
 \fi
}%
\providecommand \natexlab [1]{#1}%
\providecommand \enquote  [1]{``#1''}%
\providecommand \bibnamefont  [1]{#1}%
\providecommand \bibfnamefont [1]{#1}%
\providecommand \citenamefont [1]{#1}%
\providecommand \href@noop [0]{\@secondoftwo}%
\providecommand \href [0]{\begingroup \@sanitize@url \@href}%
\providecommand \@href[1]{\@@startlink{#1}\@@href}%
\providecommand \@@href[1]{\endgroup#1\@@endlink}%
\providecommand \@sanitize@url [0]{\catcode `\\12\catcode `\$12\catcode `\&12\catcode `\#12\catcode `\^12\catcode `\_12\catcode `\%12\relax}%
\providecommand \@@startlink[1]{}%
\providecommand \@@endlink[0]{}%
\providecommand \url  [0]{\begingroup\@sanitize@url \@url }%
\providecommand \@url [1]{\endgroup\@href {#1}{\urlprefix }}%
\providecommand \urlprefix  [0]{URL }%
\providecommand \Eprint [0]{\href }%
\providecommand \doibase [0]{https://doi.org/}%
\providecommand \selectlanguage [0]{\@gobble}%
\providecommand \bibinfo  [0]{\@secondoftwo}%
\providecommand \bibfield  [0]{\@secondoftwo}%
\providecommand \translation [1]{[#1]}%
\providecommand \BibitemOpen [0]{}%
\providecommand \bibitemStop [0]{}%
\providecommand \bibitemNoStop [0]{.\EOS\space}%
\providecommand \EOS [0]{\spacefactor3000\relax}%
\providecommand \BibitemShut  [1]{\csname bibitem#1\endcsname}%
\let\auto@bib@innerbib\@empty
\bibitem [{\citenamefont {Park}\ \emph {et~al.}(2023)\citenamefont {Park}, \citenamefont {Cai}, \citenamefont {Anderson}, \citenamefont {Zhang}, \citenamefont {Zhu}, \citenamefont {Liu}, \citenamefont {Wang}, \citenamefont {Holtzmann}, \citenamefont {Hu}, \citenamefont {Liu}, \citenamefont {Taniguchi}, \citenamefont {Watanabe}, \citenamefont {Chu}, \citenamefont {Cao}, \citenamefont {Fu}, \citenamefont {Yao}, \citenamefont {Chang}, \citenamefont {Cobden}, \citenamefont {Xiao},\ and\ \citenamefont {Xu}}]{park2023}%
  \BibitemOpen
  \bibfield  {author} {\bibinfo {author} {\bibfnamefont {H.}~\bibnamefont {Park}}, \bibinfo {author} {\bibfnamefont {J.}~\bibnamefont {Cai}}, \bibinfo {author} {\bibfnamefont {E.}~\bibnamefont {Anderson}}, \bibinfo {author} {\bibfnamefont {Y.}~\bibnamefont {Zhang}}, \bibinfo {author} {\bibfnamefont {J.}~\bibnamefont {Zhu}}, \bibinfo {author} {\bibfnamefont {X.}~\bibnamefont {Liu}}, \bibinfo {author} {\bibfnamefont {C.}~\bibnamefont {Wang}}, \bibinfo {author} {\bibfnamefont {W.}~\bibnamefont {Holtzmann}}, \bibinfo {author} {\bibfnamefont {C.}~\bibnamefont {Hu}}, \bibinfo {author} {\bibfnamefont {Z.}~\bibnamefont {Liu}}, \bibinfo {author} {\bibfnamefont {T.}~\bibnamefont {Taniguchi}}, \bibinfo {author} {\bibfnamefont {K.}~\bibnamefont {Watanabe}}, \bibinfo {author} {\bibfnamefont {J.-H.}\ \bibnamefont {Chu}}, \bibinfo {author} {\bibfnamefont {T.}~\bibnamefont {Cao}}, \bibinfo {author} {\bibfnamefont {L.}~\bibnamefont {Fu}}, \bibinfo {author} {\bibfnamefont {W.}~\bibnamefont {Yao}}, \bibinfo {author}
  {\bibfnamefont {C.-Z.}\ \bibnamefont {Chang}}, \bibinfo {author} {\bibfnamefont {D.}~\bibnamefont {Cobden}}, \bibinfo {author} {\bibfnamefont {D.}~\bibnamefont {Xiao}},\ and\ \bibinfo {author} {\bibfnamefont {X.}~\bibnamefont {Xu}},\ }\bibfield  {title} {\bibinfo {title} {Observation of fractionally quantized anomalous {{Hall}} effect},\ }\href {https://doi.org/10.1038/s41586-023-06536-0} {\bibfield  {journal} {\bibinfo  {journal} {Nature}\ }\textbf {\bibinfo {volume} {622}},\ \bibinfo {pages} {74} (\bibinfo {year} {2023})}\BibitemShut {NoStop}%
\bibitem [{\citenamefont {Cai}\ \emph {et~al.}(2023)\citenamefont {Cai}, \citenamefont {Anderson}, \citenamefont {Wang}, \citenamefont {Zhang}, \citenamefont {Liu}, \citenamefont {Holtzmann}, \citenamefont {Zhang}, \citenamefont {Fan}, \citenamefont {Taniguchi}, \citenamefont {Watanabe}, \citenamefont {Ran}, \citenamefont {Cao}, \citenamefont {Fu}, \citenamefont {Xiao}, \citenamefont {Yao},\ and\ \citenamefont {Xu}}]{cai_nature_2023}%
  \BibitemOpen
  \bibfield  {author} {\bibinfo {author} {\bibfnamefont {J.}~\bibnamefont {Cai}}, \bibinfo {author} {\bibfnamefont {E.}~\bibnamefont {Anderson}}, \bibinfo {author} {\bibfnamefont {C.}~\bibnamefont {Wang}}, \bibinfo {author} {\bibfnamefont {X.}~\bibnamefont {Zhang}}, \bibinfo {author} {\bibfnamefont {X.}~\bibnamefont {Liu}}, \bibinfo {author} {\bibfnamefont {W.}~\bibnamefont {Holtzmann}}, \bibinfo {author} {\bibfnamefont {Y.}~\bibnamefont {Zhang}}, \bibinfo {author} {\bibfnamefont {F.}~\bibnamefont {Fan}}, \bibinfo {author} {\bibfnamefont {T.}~\bibnamefont {Taniguchi}}, \bibinfo {author} {\bibfnamefont {K.}~\bibnamefont {Watanabe}}, \bibinfo {author} {\bibfnamefont {Y.}~\bibnamefont {Ran}}, \bibinfo {author} {\bibfnamefont {T.}~\bibnamefont {Cao}}, \bibinfo {author} {\bibfnamefont {L.}~\bibnamefont {Fu}}, \bibinfo {author} {\bibfnamefont {D.}~\bibnamefont {Xiao}}, \bibinfo {author} {\bibfnamefont {W.}~\bibnamefont {Yao}},\ and\ \bibinfo {author} {\bibfnamefont {X.}~\bibnamefont {Xu}},\ }\bibfield  {title}
  {\bibinfo {title} {Signatures of fractional quantum anomalous {H}all states in twisted {MoTe$_2$}},\ }\href {https://doi.org/10.1038/s41586-023-06289-w} {\bibfield  {journal} {\bibinfo  {journal} {Nature}\ }\textbf {\bibinfo {volume} {622}},\ \bibinfo {pages} {63–68} (\bibinfo {year} {2023})}\BibitemShut {NoStop}%
\bibitem [{\citenamefont {Xu}\ \emph {et~al.}(2023)\citenamefont {Xu}, \citenamefont {Sun}, \citenamefont {Jia}, \citenamefont {Liu}, \citenamefont {Xu}, \citenamefont {Li}, \citenamefont {Gu}, \citenamefont {Watanabe}, \citenamefont {Taniguchi}, \citenamefont {Tong}, \citenamefont {Jia}, \citenamefont {Shi}, \citenamefont {Jiang}, \citenamefont {Zhang}, \citenamefont {Liu},\ and\ \citenamefont {Li}}]{xu_prx_2023}%
  \BibitemOpen
  \bibfield  {author} {\bibinfo {author} {\bibfnamefont {F.}~\bibnamefont {Xu}}, \bibinfo {author} {\bibfnamefont {Z.}~\bibnamefont {Sun}}, \bibinfo {author} {\bibfnamefont {T.}~\bibnamefont {Jia}}, \bibinfo {author} {\bibfnamefont {C.}~\bibnamefont {Liu}}, \bibinfo {author} {\bibfnamefont {C.}~\bibnamefont {Xu}}, \bibinfo {author} {\bibfnamefont {C.}~\bibnamefont {Li}}, \bibinfo {author} {\bibfnamefont {Y.}~\bibnamefont {Gu}}, \bibinfo {author} {\bibfnamefont {K.}~\bibnamefont {Watanabe}}, \bibinfo {author} {\bibfnamefont {T.}~\bibnamefont {Taniguchi}}, \bibinfo {author} {\bibfnamefont {B.}~\bibnamefont {Tong}}, \bibinfo {author} {\bibfnamefont {J.}~\bibnamefont {Jia}}, \bibinfo {author} {\bibfnamefont {Z.}~\bibnamefont {Shi}}, \bibinfo {author} {\bibfnamefont {S.}~\bibnamefont {Jiang}}, \bibinfo {author} {\bibfnamefont {Y.}~\bibnamefont {Zhang}}, \bibinfo {author} {\bibfnamefont {X.}~\bibnamefont {Liu}},\ and\ \bibinfo {author} {\bibfnamefont {T.}~\bibnamefont {Li}},\ }\bibfield  {title} {\bibinfo {title}
  {Observation of integer and fractional quantum anomalous hall effects in twisted bilayer ${\mathrm{mote}}_{2}$},\ }\href {https://doi.org/10.1103/PhysRevX.13.031037} {\bibfield  {journal} {\bibinfo  {journal} {Phys. Rev. X}\ }\textbf {\bibinfo {volume} {13}},\ \bibinfo {pages} {031037} (\bibinfo {year} {2023})}\BibitemShut {NoStop}%
\bibitem [{\citenamefont {Anderson}\ \emph {et~al.}(2024)\citenamefont {Anderson}, \citenamefont {Cai}, \citenamefont {Reddy}, \citenamefont {Park}, \citenamefont {Holtzmann}, \citenamefont {Davis}, \citenamefont {Taniguchi}, \citenamefont {Watanabe}, \citenamefont {Smolenski}, \citenamefont {Imamo{\u{g}}lu}, \citenamefont {Cao}, \citenamefont {Xiao}, \citenamefont {Fu}, \citenamefont {Yao},\ and\ \citenamefont {Xu}}]{anderson_nature_2024}%
  \BibitemOpen
  \bibfield  {author} {\bibinfo {author} {\bibfnamefont {E.}~\bibnamefont {Anderson}}, \bibinfo {author} {\bibfnamefont {J.}~\bibnamefont {Cai}}, \bibinfo {author} {\bibfnamefont {A.~P.}\ \bibnamefont {Reddy}}, \bibinfo {author} {\bibfnamefont {H.}~\bibnamefont {Park}}, \bibinfo {author} {\bibfnamefont {W.}~\bibnamefont {Holtzmann}}, \bibinfo {author} {\bibfnamefont {K.}~\bibnamefont {Davis}}, \bibinfo {author} {\bibfnamefont {T.}~\bibnamefont {Taniguchi}}, \bibinfo {author} {\bibfnamefont {K.}~\bibnamefont {Watanabe}}, \bibinfo {author} {\bibfnamefont {T.}~\bibnamefont {Smolenski}}, \bibinfo {author} {\bibfnamefont {A.}~\bibnamefont {Imamo{\u{g}}lu}}, \bibinfo {author} {\bibfnamefont {T.}~\bibnamefont {Cao}}, \bibinfo {author} {\bibfnamefont {D.}~\bibnamefont {Xiao}}, \bibinfo {author} {\bibfnamefont {L.}~\bibnamefont {Fu}}, \bibinfo {author} {\bibfnamefont {W.}~\bibnamefont {Yao}},\ and\ \bibinfo {author} {\bibfnamefont {X.}~\bibnamefont {Xu}},\ }\bibfield  {title} {\bibinfo {title} {Trion sensing of a
  zero-field composite {F}ermi liquid},\ }\href {https://doi.org/10.1038/s41586-024-08134-0} {\bibfield  {journal} {\bibinfo  {journal} {Nature}\ }\textbf {\bibinfo {volume} {635}},\ \bibinfo {pages} {590} (\bibinfo {year} {2024})}\BibitemShut {NoStop}%
\bibitem [{\citenamefont {Lu}\ \emph {et~al.}(2024)\citenamefont {Lu}, \citenamefont {Han}, \citenamefont {Yao}, \citenamefont {Reddy}, \citenamefont {Yang}, \citenamefont {Seo}, \citenamefont {Watanabe}, \citenamefont {Taniguchi}, \citenamefont {Fu},\ and\ \citenamefont {Ju}}]{lu2024a}%
  \BibitemOpen
  \bibfield  {author} {\bibinfo {author} {\bibfnamefont {Z.}~\bibnamefont {Lu}}, \bibinfo {author} {\bibfnamefont {T.}~\bibnamefont {Han}}, \bibinfo {author} {\bibfnamefont {Y.}~\bibnamefont {Yao}}, \bibinfo {author} {\bibfnamefont {A.~P.}\ \bibnamefont {Reddy}}, \bibinfo {author} {\bibfnamefont {J.}~\bibnamefont {Yang}}, \bibinfo {author} {\bibfnamefont {J.}~\bibnamefont {Seo}}, \bibinfo {author} {\bibfnamefont {K.}~\bibnamefont {Watanabe}}, \bibinfo {author} {\bibfnamefont {T.}~\bibnamefont {Taniguchi}}, \bibinfo {author} {\bibfnamefont {L.}~\bibnamefont {Fu}},\ and\ \bibinfo {author} {\bibfnamefont {L.}~\bibnamefont {Ju}},\ }\bibfield  {title} {\bibinfo {title} {Fractional quantum anomalous {{Hall}} effect in multilayer graphene},\ }\href {https://doi.org/10.1038/s41586-023-07010-7} {\bibfield  {journal} {\bibinfo  {journal} {Nature}\ }\textbf {\bibinfo {volume} {626}},\ \bibinfo {pages} {759} (\bibinfo {year} {2024})}\BibitemShut {NoStop}%
\bibitem [{\citenamefont {Han}\ \emph {et~al.}(2024)\citenamefont {Han}, \citenamefont {Lu}, \citenamefont {Scuri}, \citenamefont {Sung}, \citenamefont {Wang}, \citenamefont {Han}, \citenamefont {Watanabe}, \citenamefont {Taniguchi}, \citenamefont {Park},\ and\ \citenamefont {Ju}}]{han2024a}%
  \BibitemOpen
  \bibfield  {author} {\bibinfo {author} {\bibfnamefont {T.}~\bibnamefont {Han}}, \bibinfo {author} {\bibfnamefont {Z.}~\bibnamefont {Lu}}, \bibinfo {author} {\bibfnamefont {G.}~\bibnamefont {Scuri}}, \bibinfo {author} {\bibfnamefont {J.}~\bibnamefont {Sung}}, \bibinfo {author} {\bibfnamefont {J.}~\bibnamefont {Wang}}, \bibinfo {author} {\bibfnamefont {T.}~\bibnamefont {Han}}, \bibinfo {author} {\bibfnamefont {K.}~\bibnamefont {Watanabe}}, \bibinfo {author} {\bibfnamefont {T.}~\bibnamefont {Taniguchi}}, \bibinfo {author} {\bibfnamefont {H.}~\bibnamefont {Park}},\ and\ \bibinfo {author} {\bibfnamefont {L.}~\bibnamefont {Ju}},\ }\bibfield  {title} {\bibinfo {title} {Correlated insulator and {{Chern}} insulators in pentalayer rhombohedral-stacked graphene},\ }\href {https://doi.org/10.1038/s41565-023-01520-1} {\bibfield  {journal} {\bibinfo  {journal} {Nature Nanotechnology}\ }\textbf {\bibinfo {volume} {19}},\ \bibinfo {pages} {181} (\bibinfo {year} {2024})}\BibitemShut {NoStop}%
\bibitem [{\citenamefont {Han}\ \emph {et~al.}(2025{\natexlab{a}})\citenamefont {Han}, \citenamefont {Lu}, \citenamefont {Hadjri}, \citenamefont {Shi}, \citenamefont {Wu}, \citenamefont {Xu}, \citenamefont {Yao}, \citenamefont {Cotten}, \citenamefont {Sharifi~Sedeh}, \citenamefont {Weldeyesus}, \citenamefont {Yang}, \citenamefont {Seo}, \citenamefont {Ye}, \citenamefont {Zhou}, \citenamefont {Liu}, \citenamefont {Shi}, \citenamefont {Hua}, \citenamefont {Watanabe}, \citenamefont {Taniguchi}, \citenamefont {Xiong}, \citenamefont {Zumb{\"u}hl}, \citenamefont {Fu},\ and\ \citenamefont {Ju}}]{han2025}%
  \BibitemOpen
  \bibfield  {author} {\bibinfo {author} {\bibfnamefont {T.}~\bibnamefont {Han}}, \bibinfo {author} {\bibfnamefont {Z.}~\bibnamefont {Lu}}, \bibinfo {author} {\bibfnamefont {Z.}~\bibnamefont {Hadjri}}, \bibinfo {author} {\bibfnamefont {L.}~\bibnamefont {Shi}}, \bibinfo {author} {\bibfnamefont {Z.}~\bibnamefont {Wu}}, \bibinfo {author} {\bibfnamefont {W.}~\bibnamefont {Xu}}, \bibinfo {author} {\bibfnamefont {Y.}~\bibnamefont {Yao}}, \bibinfo {author} {\bibfnamefont {A.~A.}\ \bibnamefont {Cotten}}, \bibinfo {author} {\bibfnamefont {O.}~\bibnamefont {Sharifi~Sedeh}}, \bibinfo {author} {\bibfnamefont {H.}~\bibnamefont {Weldeyesus}}, \bibinfo {author} {\bibfnamefont {J.}~\bibnamefont {Yang}}, \bibinfo {author} {\bibfnamefont {J.}~\bibnamefont {Seo}}, \bibinfo {author} {\bibfnamefont {S.}~\bibnamefont {Ye}}, \bibinfo {author} {\bibfnamefont {M.}~\bibnamefont {Zhou}}, \bibinfo {author} {\bibfnamefont {H.}~\bibnamefont {Liu}}, \bibinfo {author} {\bibfnamefont {G.}~\bibnamefont {Shi}}, \bibinfo {author} {\bibfnamefont
  {Z.}~\bibnamefont {Hua}}, \bibinfo {author} {\bibfnamefont {K.}~\bibnamefont {Watanabe}}, \bibinfo {author} {\bibfnamefont {T.}~\bibnamefont {Taniguchi}}, \bibinfo {author} {\bibfnamefont {P.}~\bibnamefont {Xiong}}, \bibinfo {author} {\bibfnamefont {D.~M.}\ \bibnamefont {Zumb{\"u}hl}}, \bibinfo {author} {\bibfnamefont {L.}~\bibnamefont {Fu}},\ and\ \bibinfo {author} {\bibfnamefont {L.}~\bibnamefont {Ju}},\ }\bibfield  {title} {\bibinfo {title} {Signatures of chiral superconductivity in rhombohedral graphene},\ }\href {https://doi.org/10.1038/s41586-025-09169-7} {\bibfield  {journal} {\bibinfo  {journal} {Nature}\ }\textbf {\bibinfo {volume} {643}},\ \bibinfo {pages} {654} (\bibinfo {year} {2025}{\natexlab{a}})}\BibitemShut {NoStop}%
\bibitem [{\citenamefont {Qin}\ \emph {et~al.}(2026)\citenamefont {Qin}, \citenamefont {Wu}, \citenamefont {Nguyen}, \citenamefont {Morissette}, \citenamefont {Zhang}, \citenamefont {Watanabe}, \citenamefont {Taniguchi},\ and\ \citenamefont {Li}}]{qin_arxiv_2026}%
  \BibitemOpen
  \bibfield  {author} {\bibinfo {author} {\bibfnamefont {P.}~\bibnamefont {Qin}}, \bibinfo {author} {\bibfnamefont {H.-T.}\ \bibnamefont {Wu}}, \bibinfo {author} {\bibfnamefont {R.~Q.}\ \bibnamefont {Nguyen}}, \bibinfo {author} {\bibfnamefont {E.}~\bibnamefont {Morissette}}, \bibinfo {author} {\bibfnamefont {N.~J.}\ \bibnamefont {Zhang}}, \bibinfo {author} {\bibfnamefont {K.}~\bibnamefont {Watanabe}}, \bibinfo {author} {\bibfnamefont {T.}~\bibnamefont {Taniguchi}},\ and\ \bibinfo {author} {\bibfnamefont {J.~I.~A.}\ \bibnamefont {Li}},\ }\href {https://arxiv.org/abs/2504.05129} {\bibinfo {title} {Extreme anisotropy in the metallic and superconducting phases of rhombohedral hexalayer graphene}} (\bibinfo {year} {2026}),\ \Eprint {https://arxiv.org/abs/2504.05129} {arXiv:2504.05129 [cond-mat.mes-hall]} \BibitemShut {NoStop}%
\bibitem [{\citenamefont {Shi}\ and\ \citenamefont {Senthil}(2025{\natexlab{a}})}]{shi_prx_2025}%
  \BibitemOpen
  \bibfield  {author} {\bibinfo {author} {\bibfnamefont {Z.~D.}\ \bibnamefont {Shi}}\ and\ \bibinfo {author} {\bibfnamefont {T.}~\bibnamefont {Senthil}},\ }\bibfield  {title} {\bibinfo {title} {{Doping a Fractional Quantum Anomalous Hall Insulator}},\ }\href {https://doi.org/10.1103/kcm5-hx56} {\bibfield  {journal} {\bibinfo  {journal} {Phys. Rev. X}\ }\textbf {\bibinfo {volume} {15}},\ \bibinfo {pages} {031069} (\bibinfo {year} {2025}{\natexlab{a}})}\BibitemShut {NoStop}%
\bibitem [{\citenamefont {Divic}\ \emph {et~al.}(2025)\citenamefont {Divic}, \citenamefont {Cr{\'e}pel}, \citenamefont {Soejima}, \citenamefont {Song}, \citenamefont {Millis}, \citenamefont {Zaletel},\ and\ \citenamefont {Vishwanath}}]{divic2025}%
  \BibitemOpen
  \bibfield  {author} {\bibinfo {author} {\bibfnamefont {S.}~\bibnamefont {Divic}}, \bibinfo {author} {\bibfnamefont {V.}~\bibnamefont {Cr{\'e}pel}}, \bibinfo {author} {\bibfnamefont {T.}~\bibnamefont {Soejima}}, \bibinfo {author} {\bibfnamefont {X.-Y.}\ \bibnamefont {Song}}, \bibinfo {author} {\bibfnamefont {A.~J.}\ \bibnamefont {Millis}}, \bibinfo {author} {\bibfnamefont {M.~P.}\ \bibnamefont {Zaletel}},\ and\ \bibinfo {author} {\bibfnamefont {A.}~\bibnamefont {Vishwanath}},\ }\bibfield  {title} {\bibinfo {title} {Anyon {{Superconductivity}} from {{Topological Criticality}} in a {{Hofstadter-Hubbard Model}}},\ }\href {https://doi.org/10.1073/pnas.2426680122} {\bibfield  {journal} {\bibinfo  {journal} {Proceedings of the National Academy of Sciences}\ }\textbf {\bibinfo {volume} {122}},\ \bibinfo {pages} {e2426680122} (\bibinfo {year} {2025})},\ \Eprint {https://arxiv.org/abs/2410.18175} {arXiv:2410.18175 [cond-mat]} \BibitemShut {NoStop}%
\bibitem [{\citenamefont {Shi}\ \emph {et~al.}(2025)\citenamefont {Shi}, \citenamefont {Zhang},\ and\ \citenamefont {Todadri}}]{shi_scipost_2025}%
  \BibitemOpen
  \bibfield  {author} {\bibinfo {author} {\bibfnamefont {Z.~D.}\ \bibnamefont {Shi}}, \bibinfo {author} {\bibfnamefont {C.}~\bibnamefont {Zhang}},\ and\ \bibinfo {author} {\bibfnamefont {S.}~\bibnamefont {Todadri}},\ }\bibfield  {title} {\bibinfo {title} {{Doping lattice non-Abelian quantum Hall states}},\ }\href {https://doi.org/10.21468/SciPostPhys.19.6.150} {\bibfield  {journal} {\bibinfo  {journal} {SciPost Phys.}\ }\textbf {\bibinfo {volume} {19}},\ \bibinfo {pages} {150} (\bibinfo {year} {2025})}\BibitemShut {NoStop}%
\bibitem [{\citenamefont {Kim}\ \emph {et~al.}(2025)\citenamefont {Kim}, \citenamefont {Timmel}, \citenamefont {Ju},\ and\ \citenamefont {Wen}}]{kim_prb_2025}%
  \BibitemOpen
  \bibfield  {author} {\bibinfo {author} {\bibfnamefont {M.}~\bibnamefont {Kim}}, \bibinfo {author} {\bibfnamefont {A.}~\bibnamefont {Timmel}}, \bibinfo {author} {\bibfnamefont {L.}~\bibnamefont {Ju}},\ and\ \bibinfo {author} {\bibfnamefont {X.-G.}\ \bibnamefont {Wen}},\ }\bibfield  {title} {\bibinfo {title} {Topological chiral superconductivity beyond pairing in a {Fermi} liquid},\ }\href {https://doi.org/10.1103/PhysRevB.111.014508} {\bibfield  {journal} {\bibinfo  {journal} {Phys. Rev. B}\ }\textbf {\bibinfo {volume} {111}},\ \bibinfo {pages} {014508} (\bibinfo {year} {2025})}\BibitemShut {NoStop}%
\bibitem [{\citenamefont {Shi}\ and\ \citenamefont {Senthil}(2025{\natexlab{b}})}]{shi_pnas_2025}%
  \BibitemOpen
  \bibfield  {author} {\bibinfo {author} {\bibfnamefont {Z.~D.}\ \bibnamefont {Shi}}\ and\ \bibinfo {author} {\bibfnamefont {T.}~\bibnamefont {Senthil}},\ }\bibfield  {title} {\bibinfo {title} {Anyon delocalization transitions out of a disordered fractional quantum anomalous {H}all insulator},\ }\href {https://doi.org/10.1073/pnas.2520608122} {\bibfield  {journal} {\bibinfo  {journal} {Proceedings of the National Academy of Sciences}\ }\textbf {\bibinfo {volume} {122}},\ \bibinfo {pages} {e2520608122} (\bibinfo {year} {2025}{\natexlab{b}})},\ \Eprint {https://arxiv.org/abs/https://www.pnas.org/doi/pdf/10.1073/pnas.2520608122} {https://www.pnas.org/doi/pdf/10.1073/pnas.2520608122} \BibitemShut {NoStop}%
\bibitem [{\citenamefont {Nosov}\ \emph {et~al.}(2026)\citenamefont {Nosov}, \citenamefont {Han},\ and\ \citenamefont {Khalaf}}]{nosov_prl_2026}%
  \BibitemOpen
  \bibfield  {author} {\bibinfo {author} {\bibfnamefont {P.~A.}\ \bibnamefont {Nosov}}, \bibinfo {author} {\bibfnamefont {Z.}~\bibnamefont {Han}},\ and\ \bibinfo {author} {\bibfnamefont {E.}~\bibnamefont {Khalaf}},\ }\bibfield  {title} {\bibinfo {title} {Anyon superconductivity and plateau transitions in doped fractional quantum anomalous hall insulators},\ }\href {https://doi.org/10.1103/6bgj-bfdn} {\bibfield  {journal} {\bibinfo  {journal} {Phys. Rev. Lett.}\ }\textbf {\bibinfo {volume} {136}},\ \bibinfo {pages} {106501} (\bibinfo {year} {2026})}\BibitemShut {NoStop}%
\bibitem [{\citenamefont {Pichler}\ \emph {et~al.}(2025)\citenamefont {Pichler}, \citenamefont {Kuhlenkamp}, \citenamefont {Knap},\ and\ \citenamefont {Vishwanath}}]{pichler2025arxiv}%
  \BibitemOpen
  \bibfield  {author} {\bibinfo {author} {\bibfnamefont {F.}~\bibnamefont {Pichler}}, \bibinfo {author} {\bibfnamefont {C.}~\bibnamefont {Kuhlenkamp}}, \bibinfo {author} {\bibfnamefont {M.}~\bibnamefont {Knap}},\ and\ \bibinfo {author} {\bibfnamefont {A.}~\bibnamefont {Vishwanath}},\ }\href {https://arxiv.org/abs/2506.08000} {\bibinfo {title} {Microscopic {M}echanism of {A}nyon {S}uperconductivity {E}merging from {F}ractional {C}hern {I}nsulators}} (\bibinfo {year} {2025}),\ \Eprint {https://arxiv.org/abs/2506.08000} {arXiv:2506.08000 [cond-mat.str-el]} \BibitemShut {NoStop}%
\bibitem [{\citenamefont {Wang}\ and\ \citenamefont {Zaletel}(2025)}]{wang2025chiral}%
  \BibitemOpen
  \bibfield  {author} {\bibinfo {author} {\bibfnamefont {T.}~\bibnamefont {Wang}}\ and\ \bibinfo {author} {\bibfnamefont {M.~P.}\ \bibnamefont {Zaletel}},\ }\href {https://arxiv.org/abs/2507.07921} {\bibinfo {title} {Chiral superconductivity near a fractional {C}hern insulator}} (\bibinfo {year} {2025}),\ \Eprint {https://arxiv.org/abs/2507.07921} {arXiv:2507.07921 [cond-mat.str-el]} \BibitemShut {NoStop}%
\bibitem [{\citenamefont {Han}\ \emph {et~al.}(2025{\natexlab{b}})\citenamefont {Han}, \citenamefont {Wang}, \citenamefont {Dong}, \citenamefont {Zaletel},\ and\ \citenamefont {Vishwanath}}]{han2025a}%
  \BibitemOpen
  \bibfield  {author} {\bibinfo {author} {\bibfnamefont {Z.}~\bibnamefont {Han}}, \bibinfo {author} {\bibfnamefont {T.}~\bibnamefont {Wang}}, \bibinfo {author} {\bibfnamefont {Z.}~\bibnamefont {Dong}}, \bibinfo {author} {\bibfnamefont {M.~P.}\ \bibnamefont {Zaletel}},\ and\ \bibinfo {author} {\bibfnamefont {A.}~\bibnamefont {Vishwanath}},\ }\href {https://doi.org/10.48550/arXiv.2508.14894} {\bibinfo {title} {Anyon superfluidity of excitons in quantum {{Hall}} bilayers}} (\bibinfo {year} {2025}{\natexlab{b}}),\ \Eprint {https://arxiv.org/abs/2508.14894} {arXiv:2508.14894 [cond-mat]} \BibitemShut {NoStop}%
\bibitem [{\citenamefont {Wang}\ and\ \citenamefont {Zhang}(2025)}]{wang2025}%
  \BibitemOpen
  \bibfield  {author} {\bibinfo {author} {\bibfnamefont {T.}~\bibnamefont {Wang}}\ and\ \bibinfo {author} {\bibfnamefont {Y.-H.}\ \bibnamefont {Zhang}},\ }\href {https://doi.org/10.48550/arXiv.2508.00058} {\bibinfo {title} {Anyon superfluid in trilayer quantum {{Hall}} systems}} (\bibinfo {year} {2025}),\ \Eprint {https://arxiv.org/abs/2508.00058} {arXiv:2508.00058 [cond-mat]} \BibitemShut {NoStop}%
\bibitem [{Note1()}]{Note1}%
  \BibitemOpen
  \bibinfo {note} {It may be pertinent to the superconductivity reported in Ref.~\cite {xu2025arxiv} in tMoTe$_2$ close to the 2/3 FQAH state (see Refs.~\cite {shi_prx_2025,shi_pnas_2025,nosov_prl_2026}).}\BibitemShut {Stop}%
\bibitem [{\citenamefont {Halperin}\ \emph {et~al.}(1993)\citenamefont {Halperin}, \citenamefont {Lee},\ and\ \citenamefont {Read}}]{halperin1993}%
  \BibitemOpen
  \bibfield  {author} {\bibinfo {author} {\bibfnamefont {B.~I.}\ \bibnamefont {Halperin}}, \bibinfo {author} {\bibfnamefont {P.~A.}\ \bibnamefont {Lee}},\ and\ \bibinfo {author} {\bibfnamefont {N.}~\bibnamefont {Read}},\ }\bibfield  {title} {\bibinfo {title} {Theory of the half-filled {{Landau}} level},\ }\href {https://doi.org/10.1103/PhysRevB.47.7312} {\bibfield  {journal} {\bibinfo  {journal} {Physical Review B}\ }\textbf {\bibinfo {volume} {47}},\ \bibinfo {pages} {7312} (\bibinfo {year} {1993})}\BibitemShut {NoStop}%
\bibitem [{\citenamefont {Barkeshli}\ and\ \citenamefont {McGreevy}(2012)}]{barkeshli_prb_2012}%
  \BibitemOpen
  \bibfield  {author} {\bibinfo {author} {\bibfnamefont {M.}~\bibnamefont {Barkeshli}}\ and\ \bibinfo {author} {\bibfnamefont {J.}~\bibnamefont {McGreevy}},\ }\bibfield  {title} {\bibinfo {title} {{Continuous transitions between composite Fermi liquid and Landau Fermi liquid: A route to fractionalized Mott insulators}},\ }\href {https://doi.org/10.1103/PhysRevB.86.075136} {\bibfield  {journal} {\bibinfo  {journal} {Phys. Rev. B}\ }\textbf {\bibinfo {volume} {86}},\ \bibinfo {pages} {075136} (\bibinfo {year} {2012})}\BibitemShut {NoStop}%
\bibitem [{\citenamefont {Song}\ \emph {et~al.}(2024)\citenamefont {Song}, \citenamefont {Zhang},\ and\ \citenamefont {Senthil}}]{song_prb_2024}%
  \BibitemOpen
  \bibfield  {author} {\bibinfo {author} {\bibfnamefont {X.-Y.}\ \bibnamefont {Song}}, \bibinfo {author} {\bibfnamefont {Y.-H.}\ \bibnamefont {Zhang}},\ and\ \bibinfo {author} {\bibfnamefont {T.}~\bibnamefont {Senthil}},\ }\bibfield  {title} {\bibinfo {title} {{Phase transitions out of quantum Hall states in moir\'e materials}},\ }\href {https://doi.org/10.1103/PhysRevB.109.085143} {\bibfield  {journal} {\bibinfo  {journal} {Phys. Rev. B}\ }\textbf {\bibinfo {volume} {109}},\ \bibinfo {pages} {085143} (\bibinfo {year} {2024})}\BibitemShut {NoStop}%
\bibitem [{Note2()}]{Note2}%
  \BibitemOpen
  \bibinfo {note} {It is possible that it originates from what microscopically is a repulsive interaction through the Kohn-Luttinger mechanism, or through some other exotic route \cite {kim_prb_2025}.}\BibitemShut {Stop}%
\bibitem [{\citenamefont {Guerci}\ \emph {et~al.}(2024)\citenamefont {Guerci}, \citenamefont {Kaplan}, \citenamefont {Ingham}, \citenamefont {Pixley},\ and\ \citenamefont {Millis}}]{guerci2024arxiv}%
  \BibitemOpen
  \bibfield  {author} {\bibinfo {author} {\bibfnamefont {D.}~\bibnamefont {Guerci}}, \bibinfo {author} {\bibfnamefont {D.}~\bibnamefont {Kaplan}}, \bibinfo {author} {\bibfnamefont {J.}~\bibnamefont {Ingham}}, \bibinfo {author} {\bibfnamefont {J.~H.}\ \bibnamefont {Pixley}},\ and\ \bibinfo {author} {\bibfnamefont {A.~J.}\ \bibnamefont {Millis}},\ }\href {https://arxiv.org/abs/2408.16075} {\bibinfo {title} {Topological superconductivity from repulsive interactions in twisted {WS}e$_2$}} (\bibinfo {year} {2024}),\ \Eprint {https://arxiv.org/abs/2408.16075} {arXiv:2408.16075 [cond-mat.supr-con]} \BibitemShut {NoStop}%
\bibitem [{\citenamefont {Chou}\ \emph {et~al.}(2025)\citenamefont {Chou}, \citenamefont {Zhu},\ and\ \citenamefont {Das~Sarma}}]{chou_prb_2025}%
  \BibitemOpen
  \bibfield  {author} {\bibinfo {author} {\bibfnamefont {Y.-Z.}\ \bibnamefont {Chou}}, \bibinfo {author} {\bibfnamefont {J.}~\bibnamefont {Zhu}},\ and\ \bibinfo {author} {\bibfnamefont {S.}~\bibnamefont {Das~Sarma}},\ }\bibfield  {title} {\bibinfo {title} {Intravalley spin-polarized superconductivity in rhombohedral tetralayer graphene},\ }\href {https://doi.org/10.1103/PhysRevB.111.174523} {\bibfield  {journal} {\bibinfo  {journal} {Phys. Rev. B}\ }\textbf {\bibinfo {volume} {111}},\ \bibinfo {pages} {174523} (\bibinfo {year} {2025})}\BibitemShut {NoStop}%
\bibitem [{\citenamefont {Qin}\ \emph {et~al.}(2025)\citenamefont {Qin}, \citenamefont {Qiu},\ and\ \citenamefont {Wu}}]{qin_prl_2025}%
  \BibitemOpen
  \bibfield  {author} {\bibinfo {author} {\bibfnamefont {W.}~\bibnamefont {Qin}}, \bibinfo {author} {\bibfnamefont {W.-X.}\ \bibnamefont {Qiu}},\ and\ \bibinfo {author} {\bibfnamefont {F.}~\bibnamefont {Wu}},\ }\bibfield  {title} {\bibinfo {title} {{Topological Chiral Superconductivity Mediated by Intervalley Antiferromagnetic Fluctuations in Twisted Bilayer ${\mathrm{WSe}}_{2}$}},\ }\href {https://doi.org/10.1103/kf2b-r9g5} {\bibfield  {journal} {\bibinfo  {journal} {Phys. Rev. Lett.}\ }\textbf {\bibinfo {volume} {135}},\ \bibinfo {pages} {246002} (\bibinfo {year} {2025})}\BibitemShut {NoStop}%
\bibitem [{\citenamefont {Jahin}\ and\ \citenamefont {Lin}(2026)}]{jahin_prb_2026}%
  \BibitemOpen
  \bibfield  {author} {\bibinfo {author} {\bibfnamefont {A.}~\bibnamefont {Jahin}}\ and\ \bibinfo {author} {\bibfnamefont {S.-Z.}\ \bibnamefont {Lin}},\ }\bibfield  {title} {\bibinfo {title} {{Enhanced Kohn-Luttinger superconductivity in geometric bands}},\ }\href {https://doi.org/10.1103/gt8h-czf3} {\bibfield  {journal} {\bibinfo  {journal} {Phys. Rev. B}\ }\textbf {\bibinfo {volume} {113}},\ \bibinfo {pages} {014504} (\bibinfo {year} {2026})}\BibitemShut {NoStop}%
\bibitem [{\citenamefont {Xu}\ \emph {et~al.}(2025{\natexlab{a}})\citenamefont {Xu}, \citenamefont {Zou}, \citenamefont {Peshcherenko}, \citenamefont {Jahin}, \citenamefont {Li}, \citenamefont {Lin},\ and\ \citenamefont {Zhang}}]{xu_prl_2025}%
  \BibitemOpen
  \bibfield  {author} {\bibinfo {author} {\bibfnamefont {C.}~\bibnamefont {Xu}}, \bibinfo {author} {\bibfnamefont {N.}~\bibnamefont {Zou}}, \bibinfo {author} {\bibfnamefont {N.}~\bibnamefont {Peshcherenko}}, \bibinfo {author} {\bibfnamefont {A.}~\bibnamefont {Jahin}}, \bibinfo {author} {\bibfnamefont {T.}~\bibnamefont {Li}}, \bibinfo {author} {\bibfnamefont {S.-Z.}\ \bibnamefont {Lin}},\ and\ \bibinfo {author} {\bibfnamefont {Y.}~\bibnamefont {Zhang}},\ }\bibfield  {title} {\bibinfo {title} {{Chiral Superconductivity from Spin Polarized Chern Band in Twisted ${\mathrm{MoTe}}_{2}$}},\ }\href {https://doi.org/10.1103/h22z-4hsj} {\bibfield  {journal} {\bibinfo  {journal} {Phys. Rev. Lett.}\ }\textbf {\bibinfo {volume} {135}},\ \bibinfo {pages} {266005} (\bibinfo {year} {2025}{\natexlab{a}})}\BibitemShut {NoStop}%
\bibitem [{\citenamefont {Guerci}\ \emph {et~al.}(2025)\citenamefont {Guerci}, \citenamefont {Abouelkomsan},\ and\ \citenamefont {Fu}}]{guerci_prl_2025}%
  \BibitemOpen
  \bibfield  {author} {\bibinfo {author} {\bibfnamefont {D.}~\bibnamefont {Guerci}}, \bibinfo {author} {\bibfnamefont {A.}~\bibnamefont {Abouelkomsan}},\ and\ \bibinfo {author} {\bibfnamefont {L.}~\bibnamefont {Fu}},\ }\bibfield  {title} {\bibinfo {title} {{From Fractionalization to Chiral Topological Superconductivity in a Flat Chern Band}},\ }\href {https://doi.org/10.1103/zm39-dstj} {\bibfield  {journal} {\bibinfo  {journal} {Phys. Rev. Lett.}\ }\textbf {\bibinfo {volume} {135}},\ \bibinfo {pages} {186601} (\bibinfo {year} {2025})}\BibitemShut {NoStop}%
\bibitem [{\citenamefont {Chen}\ \emph {et~al.}(2026)\citenamefont {Chen}, \citenamefont {Xu}, \citenamefont {Zhang},\ and\ \citenamefont {Schrade}}]{chen_nat_2026}%
  \BibitemOpen
  \bibfield  {author} {\bibinfo {author} {\bibfnamefont {Y.}~\bibnamefont {Chen}}, \bibinfo {author} {\bibfnamefont {C.}~\bibnamefont {Xu}}, \bibinfo {author} {\bibfnamefont {Y.}~\bibnamefont {Zhang}},\ and\ \bibinfo {author} {\bibfnamefont {C.}~\bibnamefont {Schrade}},\ }\bibfield  {title} {\bibinfo {title} {{Finite-momentum superconductivity from chiral bands in twisted MoTe$_2$}},\ }\href {https://doi.org/10.1038/s41467-025-67836-9} {\bibfield  {journal} {\bibinfo  {journal} {Nature Communications}\ } (\bibinfo {year} {2026})}\BibitemShut {NoStop}%
\bibitem [{\citenamefont {Shayegan}(2020)}]{shayegan2020}%
  \BibitemOpen
  \bibfield  {author} {\bibinfo {author} {\bibfnamefont {M.}~\bibnamefont {Shayegan}},\ }\bibfield  {title} {\bibinfo {title} {Probing {{Composite Fermions Near Half-Filled Landau Levels}}},\ }in\ \href {https://doi.org/10.1142/9789811217494_0003} {\emph {\bibinfo {booktitle} {Fractional {{Quantum Hall Effects}}}}}\ (\bibinfo  {publisher} {WORLD SCIENTIFIC},\ \bibinfo {year} {2020})\ pp.\ \bibinfo {pages} {133--181}\BibitemShut {NoStop}%
\bibitem [{\citenamefont {Eisenstein}\ \emph {et~al.}(1992)\citenamefont {Eisenstein}, \citenamefont {Pfeiffer},\ and\ \citenamefont {West}}]{eisenstein1992}%
  \BibitemOpen
  \bibfield  {author} {\bibinfo {author} {\bibfnamefont {J.~P.}\ \bibnamefont {Eisenstein}}, \bibinfo {author} {\bibfnamefont {L.~N.}\ \bibnamefont {Pfeiffer}},\ and\ \bibinfo {author} {\bibfnamefont {K.~W.}\ \bibnamefont {West}},\ }\bibfield  {title} {\bibinfo {title} {Coulomb barrier to tunneling between parallel two-dimensional electron systems},\ }\href {https://doi.org/10.1103/PhysRevLett.69.3804} {\bibfield  {journal} {\bibinfo  {journal} {Physical Review Letters}\ }\textbf {\bibinfo {volume} {69}},\ \bibinfo {pages} {3804} (\bibinfo {year} {1992})}\BibitemShut {NoStop}%
\bibitem [{\citenamefont {Willett}\ \emph {et~al.}(1993{\natexlab{a}})\citenamefont {Willett}, \citenamefont {Ruel}, \citenamefont {West},\ and\ \citenamefont {Pfeiffer}}]{willett1993}%
  \BibitemOpen
  \bibfield  {author} {\bibinfo {author} {\bibfnamefont {R.~L.}\ \bibnamefont {Willett}}, \bibinfo {author} {\bibfnamefont {R.~R.}\ \bibnamefont {Ruel}}, \bibinfo {author} {\bibfnamefont {K.~W.}\ \bibnamefont {West}},\ and\ \bibinfo {author} {\bibfnamefont {L.~N.}\ \bibnamefont {Pfeiffer}},\ }\bibfield  {title} {\bibinfo {title} {Experimental demonstration of a {{Fermi}} surface at one-half filling of the lowest {{Landau}} level},\ }\href {https://doi.org/10.1103/PhysRevLett.71.3846} {\bibfield  {journal} {\bibinfo  {journal} {Physical Review Letters}\ }\textbf {\bibinfo {volume} {71}},\ \bibinfo {pages} {3846} (\bibinfo {year} {1993}{\natexlab{a}})}\BibitemShut {NoStop}%
\bibitem [{\citenamefont {Kang}\ \emph {et~al.}(1993)\citenamefont {Kang}, \citenamefont {Stormer}, \citenamefont {Pfeiffer}, \citenamefont {Baldwin},\ and\ \citenamefont {West}}]{kang1993}%
  \BibitemOpen
  \bibfield  {author} {\bibinfo {author} {\bibfnamefont {W.}~\bibnamefont {Kang}}, \bibinfo {author} {\bibfnamefont {H.~L.}\ \bibnamefont {Stormer}}, \bibinfo {author} {\bibfnamefont {L.~N.}\ \bibnamefont {Pfeiffer}}, \bibinfo {author} {\bibfnamefont {K.~W.}\ \bibnamefont {Baldwin}},\ and\ \bibinfo {author} {\bibfnamefont {K.~W.}\ \bibnamefont {West}},\ }\bibfield  {title} {\bibinfo {title} {How real are composite fermions?},\ }\href {https://doi.org/10.1103/PhysRevLett.71.3850} {\bibfield  {journal} {\bibinfo  {journal} {Physical Review Letters}\ }\textbf {\bibinfo {volume} {71}},\ \bibinfo {pages} {3850} (\bibinfo {year} {1993})}\BibitemShut {NoStop}%
\bibitem [{\citenamefont {Willett}\ \emph {et~al.}(1993{\natexlab{b}})\citenamefont {Willett}, \citenamefont {Ruel}, \citenamefont {Paalanen}, \citenamefont {West},\ and\ \citenamefont {Pfeiffer}}]{willett1993a}%
  \BibitemOpen
  \bibfield  {author} {\bibinfo {author} {\bibfnamefont {R.~L.}\ \bibnamefont {Willett}}, \bibinfo {author} {\bibfnamefont {R.~R.}\ \bibnamefont {Ruel}}, \bibinfo {author} {\bibfnamefont {M.~A.}\ \bibnamefont {Paalanen}}, \bibinfo {author} {\bibfnamefont {K.~W.}\ \bibnamefont {West}},\ and\ \bibinfo {author} {\bibfnamefont {L.~N.}\ \bibnamefont {Pfeiffer}},\ }\bibfield  {title} {\bibinfo {title} {Enhanced finite-wave-vector conductivity at multiple even-denominator filling factors in two-dimensional electron systems},\ }\href {https://doi.org/10.1103/PhysRevB.47.7344} {\bibfield  {journal} {\bibinfo  {journal} {Physical Review B}\ }\textbf {\bibinfo {volume} {47}},\ \bibinfo {pages} {7344} (\bibinfo {year} {1993}{\natexlab{b}})}\BibitemShut {NoStop}%
\bibitem [{\citenamefont {Goldman}\ \emph {et~al.}(1994)\citenamefont {Goldman}, \citenamefont {Su},\ and\ \citenamefont {Jain}}]{goldman1994}%
  \BibitemOpen
  \bibfield  {author} {\bibinfo {author} {\bibfnamefont {V.~J.}\ \bibnamefont {Goldman}}, \bibinfo {author} {\bibfnamefont {B.}~\bibnamefont {Su}},\ and\ \bibinfo {author} {\bibfnamefont {J.~K.}\ \bibnamefont {Jain}},\ }\bibfield  {title} {\bibinfo {title} {Detection of composite fermions by magnetic focusing},\ }\href {https://doi.org/10.1103/PhysRevLett.72.2065} {\bibfield  {journal} {\bibinfo  {journal} {Physical Review Letters}\ }\textbf {\bibinfo {volume} {72}},\ \bibinfo {pages} {2065} (\bibinfo {year} {1994})}\BibitemShut {NoStop}%
\bibitem [{\citenamefont {Smet}\ \emph {et~al.}(1996)\citenamefont {Smet}, \citenamefont {Weiss}, \citenamefont {Blick}, \citenamefont {L{\"u}tjering}, \citenamefont {{von Klitzing}}, \citenamefont {Fleischmann}, \citenamefont {Ketzmerick}, \citenamefont {Geisel},\ and\ \citenamefont {Weimann}}]{smet1996}%
  \BibitemOpen
  \bibfield  {author} {\bibinfo {author} {\bibfnamefont {J.~H.}\ \bibnamefont {Smet}}, \bibinfo {author} {\bibfnamefont {D.}~\bibnamefont {Weiss}}, \bibinfo {author} {\bibfnamefont {R.~H.}\ \bibnamefont {Blick}}, \bibinfo {author} {\bibfnamefont {G.}~\bibnamefont {L{\"u}tjering}}, \bibinfo {author} {\bibfnamefont {K.}~\bibnamefont {{von Klitzing}}}, \bibinfo {author} {\bibfnamefont {R.}~\bibnamefont {Fleischmann}}, \bibinfo {author} {\bibfnamefont {R.}~\bibnamefont {Ketzmerick}}, \bibinfo {author} {\bibfnamefont {T.}~\bibnamefont {Geisel}},\ and\ \bibinfo {author} {\bibfnamefont {G.}~\bibnamefont {Weimann}},\ }\bibfield  {title} {\bibinfo {title} {Magnetic {{Focusing}} of {{Composite Fermions}} through {{Arrays}} of {{Cavities}}},\ }\href {https://doi.org/10.1103/PhysRevLett.77.2272} {\bibfield  {journal} {\bibinfo  {journal} {Physical Review Letters}\ }\textbf {\bibinfo {volume} {77}},\ \bibinfo {pages} {2272} (\bibinfo {year} {1996})}\BibitemShut {NoStop}%
\bibitem [{\citenamefont {Dong}\ \emph {et~al.}(2023)\citenamefont {Dong}, \citenamefont {Wang}, \citenamefont {Ledwith}, \citenamefont {Vishwanath},\ and\ \citenamefont {Parker}}]{dong2023a}%
  \BibitemOpen
  \bibfield  {author} {\bibinfo {author} {\bibfnamefont {J.}~\bibnamefont {Dong}}, \bibinfo {author} {\bibfnamefont {J.}~\bibnamefont {Wang}}, \bibinfo {author} {\bibfnamefont {P.~J.}\ \bibnamefont {Ledwith}}, \bibinfo {author} {\bibfnamefont {A.}~\bibnamefont {Vishwanath}},\ and\ \bibinfo {author} {\bibfnamefont {D.~E.}\ \bibnamefont {Parker}},\ }\bibfield  {title} {\bibinfo {title} {Composite {{Fermi Liquid}} at {{Zero Magnetic Field}} in {{Twisted}} \$\textbraceleft\textbackslash mathrm\textbraceleft{{MoTe}}\textbraceright\textbraceright\_\textbraceleft 2\textbraceright\$},\ }\href {https://doi.org/10.1103/PhysRevLett.131.136502} {\bibfield  {journal} {\bibinfo  {journal} {Physical Review Letters}\ }\textbf {\bibinfo {volume} {131}},\ \bibinfo {pages} {136502} (\bibinfo {year} {2023})}\BibitemShut {NoStop}%
\bibitem [{\citenamefont {Goldman}\ \emph {et~al.}(2023)\citenamefont {Goldman}, \citenamefont {Reddy}, \citenamefont {Paul},\ and\ \citenamefont {Fu}}]{goldman2023}%
  \BibitemOpen
  \bibfield  {author} {\bibinfo {author} {\bibfnamefont {H.}~\bibnamefont {Goldman}}, \bibinfo {author} {\bibfnamefont {A.~P.}\ \bibnamefont {Reddy}}, \bibinfo {author} {\bibfnamefont {N.}~\bibnamefont {Paul}},\ and\ \bibinfo {author} {\bibfnamefont {L.}~\bibnamefont {Fu}},\ }\bibfield  {title} {\bibinfo {title} {Zero-{{Field Composite Fermi Liquid}} in {{Twisted Semiconductor Bilayers}}},\ }\href {https://doi.org/10.1103/PhysRevLett.131.136501} {\bibfield  {journal} {\bibinfo  {journal} {Physical Review Letters}\ }\textbf {\bibinfo {volume} {131}},\ \bibinfo {pages} {136501} (\bibinfo {year} {2023})}\BibitemShut {NoStop}%
\bibitem [{\citenamefont {Lu}\ \emph {et~al.}(2025)\citenamefont {Lu}, \citenamefont {Han}, \citenamefont {Yao}, \citenamefont {Hadjri}, \citenamefont {Yang}, \citenamefont {Seo}, \citenamefont {Shi}, \citenamefont {Ye}, \citenamefont {Watanabe}, \citenamefont {Taniguchi},\ and\ \citenamefont {Ju}}]{lu_nature_2025}%
  \BibitemOpen
  \bibfield  {author} {\bibinfo {author} {\bibfnamefont {Z.}~\bibnamefont {Lu}}, \bibinfo {author} {\bibfnamefont {T.}~\bibnamefont {Han}}, \bibinfo {author} {\bibfnamefont {Y.}~\bibnamefont {Yao}}, \bibinfo {author} {\bibfnamefont {Z.}~\bibnamefont {Hadjri}}, \bibinfo {author} {\bibfnamefont {J.}~\bibnamefont {Yang}}, \bibinfo {author} {\bibfnamefont {J.}~\bibnamefont {Seo}}, \bibinfo {author} {\bibfnamefont {L.}~\bibnamefont {Shi}}, \bibinfo {author} {\bibfnamefont {S.}~\bibnamefont {Ye}}, \bibinfo {author} {\bibfnamefont {K.}~\bibnamefont {Watanabe}}, \bibinfo {author} {\bibfnamefont {T.}~\bibnamefont {Taniguchi}},\ and\ \bibinfo {author} {\bibfnamefont {L.}~\bibnamefont {Ju}},\ }\bibfield  {title} {\bibinfo {title} {Extended quantum anomalous hall states in {graphene/hBN} moir{\'e} superlattices},\ }\href {https://doi.org/10.1038/s41586-024-08470-1} {\bibfield  {journal} {\bibinfo  {journal} {Nature}\ }\textbf {\bibinfo {volume} {637}},\ \bibinfo {pages} {1090} (\bibinfo {year} {2025})}\BibitemShut
  {NoStop}%
\bibitem [{\citenamefont {Metlitski}\ \emph {et~al.}(2015)\citenamefont {Metlitski}, \citenamefont {Mross}, \citenamefont {Sachdev},\ and\ \citenamefont {Senthil}}]{metlitski2015a}%
  \BibitemOpen
  \bibfield  {author} {\bibinfo {author} {\bibfnamefont {M.~A.}\ \bibnamefont {Metlitski}}, \bibinfo {author} {\bibfnamefont {D.~F.}\ \bibnamefont {Mross}}, \bibinfo {author} {\bibfnamefont {S.}~\bibnamefont {Sachdev}},\ and\ \bibinfo {author} {\bibfnamefont {T.}~\bibnamefont {Senthil}},\ }\bibfield  {title} {\bibinfo {title} {Cooper pairing in non-{{Fermi}} liquids},\ }\href {https://doi.org/10.1103/PhysRevB.91.115111} {\bibfield  {journal} {\bibinfo  {journal} {Physical Review B}\ }\textbf {\bibinfo {volume} {91}},\ \bibinfo {pages} {115111} (\bibinfo {year} {2015})}\BibitemShut {NoStop}%
\bibitem [{\citenamefont {Moore}\ and\ \citenamefont {Read}(1991)}]{moore1991}%
  \BibitemOpen
  \bibfield  {author} {\bibinfo {author} {\bibfnamefont {G.}~\bibnamefont {Moore}}\ and\ \bibinfo {author} {\bibfnamefont {N.}~\bibnamefont {Read}},\ }\bibfield  {title} {\bibinfo {title} {Nonabelions in the fractional quantum hall effect},\ }\href {https://doi.org/10.1016/0550-3213(91)90407-O} {\bibfield  {journal} {\bibinfo  {journal} {Nuclear Physics B}\ }\textbf {\bibinfo {volume} {360}},\ \bibinfo {pages} {362} (\bibinfo {year} {1991})}\BibitemShut {NoStop}%
\bibitem [{\citenamefont {Read}\ and\ \citenamefont {Green}(2000)}]{read2000}%
  \BibitemOpen
  \bibfield  {author} {\bibinfo {author} {\bibfnamefont {N.}~\bibnamefont {Read}}\ and\ \bibinfo {author} {\bibfnamefont {D.}~\bibnamefont {Green}},\ }\bibfield  {title} {\bibinfo {title} {Paired states of fermions in two dimensions with breaking of parity and time-reversal symmetries and the fractional quantum {{Hall}} effect},\ }\href {https://doi.org/10.1103/PhysRevB.61.10267} {\bibfield  {journal} {\bibinfo  {journal} {Physical Review B}\ }\textbf {\bibinfo {volume} {61}},\ \bibinfo {pages} {10267} (\bibinfo {year} {2000})}\BibitemShut {NoStop}%
\bibitem [{\citenamefont {Willett}\ \emph {et~al.}(1987)\citenamefont {Willett}, \citenamefont {Eisenstein}, \citenamefont {St{\"o}rmer}, \citenamefont {Tsui}, \citenamefont {Gossard},\ and\ \citenamefont {English}}]{willett1987}%
  \BibitemOpen
  \bibfield  {author} {\bibinfo {author} {\bibfnamefont {R.}~\bibnamefont {Willett}}, \bibinfo {author} {\bibfnamefont {J.~P.}\ \bibnamefont {Eisenstein}}, \bibinfo {author} {\bibfnamefont {H.~L.}\ \bibnamefont {St{\"o}rmer}}, \bibinfo {author} {\bibfnamefont {D.~C.}\ \bibnamefont {Tsui}}, \bibinfo {author} {\bibfnamefont {A.~C.}\ \bibnamefont {Gossard}},\ and\ \bibinfo {author} {\bibfnamefont {J.~H.}\ \bibnamefont {English}},\ }\bibfield  {title} {\bibinfo {title} {Observation of an even-denominator quantum number in the fractional quantum {{Hall}} effect},\ }\href {https://doi.org/10.1103/PhysRevLett.59.1776} {\bibfield  {journal} {\bibinfo  {journal} {Physical Review Letters}\ }\textbf {\bibinfo {volume} {59}},\ \bibinfo {pages} {1776} (\bibinfo {year} {1987})}\BibitemShut {NoStop}%
\bibitem [{\citenamefont {Radu}\ \emph {et~al.}(2008)\citenamefont {Radu}, \citenamefont {Miller}, \citenamefont {Marcus}, \citenamefont {Kastner}, \citenamefont {Pfeiffer},\ and\ \citenamefont {West}}]{radu2008}%
  \BibitemOpen
  \bibfield  {author} {\bibinfo {author} {\bibfnamefont {I.~P.}\ \bibnamefont {Radu}}, \bibinfo {author} {\bibfnamefont {J.~B.}\ \bibnamefont {Miller}}, \bibinfo {author} {\bibfnamefont {C.~M.}\ \bibnamefont {Marcus}}, \bibinfo {author} {\bibfnamefont {M.~A.}\ \bibnamefont {Kastner}}, \bibinfo {author} {\bibfnamefont {L.~N.}\ \bibnamefont {Pfeiffer}},\ and\ \bibinfo {author} {\bibfnamefont {K.~W.}\ \bibnamefont {West}},\ }\bibfield  {title} {\bibinfo {title} {Quasi-{{Particle Properties}} from {{Tunneling}} in the v = 5/2 {{Fractional Quantum Hall State}}},\ }\href {https://doi.org/10.1126/science.1157560} {\bibfield  {journal} {\bibinfo  {journal} {Science}\ }\textbf {\bibinfo {volume} {320}},\ \bibinfo {pages} {899} (\bibinfo {year} {2008})}\BibitemShut {NoStop}%
\bibitem [{\citenamefont {Dolev}\ \emph {et~al.}(2008)\citenamefont {Dolev}, \citenamefont {Heiblum}, \citenamefont {Umansky}, \citenamefont {Stern},\ and\ \citenamefont {Mahalu}}]{dolev2008}%
  \BibitemOpen
  \bibfield  {author} {\bibinfo {author} {\bibfnamefont {M.}~\bibnamefont {Dolev}}, \bibinfo {author} {\bibfnamefont {M.}~\bibnamefont {Heiblum}}, \bibinfo {author} {\bibfnamefont {V.}~\bibnamefont {Umansky}}, \bibinfo {author} {\bibfnamefont {A.}~\bibnamefont {Stern}},\ and\ \bibinfo {author} {\bibfnamefont {D.}~\bibnamefont {Mahalu}},\ }\bibfield  {title} {\bibinfo {title} {Observation of a quarter of an electron charge at the {$\nu$} = 5/2 quantum {{Hall}} state},\ }\href {https://doi.org/10.1038/nature06855} {\bibfield  {journal} {\bibinfo  {journal} {Nature}\ }\textbf {\bibinfo {volume} {452}},\ \bibinfo {pages} {829} (\bibinfo {year} {2008})}\BibitemShut {NoStop}%
\bibitem [{\citenamefont {Storni}\ \emph {et~al.}(2010)\citenamefont {Storni}, \citenamefont {Morf},\ and\ \citenamefont {Das~Sarma}}]{storni2010}%
  \BibitemOpen
  \bibfield  {author} {\bibinfo {author} {\bibfnamefont {M.}~\bibnamefont {Storni}}, \bibinfo {author} {\bibfnamefont {R.~H.}\ \bibnamefont {Morf}},\ and\ \bibinfo {author} {\bibfnamefont {S.}~\bibnamefont {Das~Sarma}},\ }\bibfield  {title} {\bibinfo {title} {Fractional {{Quantum Hall State}} at {$\nu=5/2$} and the {{Moore-Read Pfaffian}}},\ }\href {https://doi.org/10.1103/PhysRevLett.104.076803} {\bibfield  {journal} {\bibinfo  {journal} {Physical Review Letters}\ }\textbf {\bibinfo {volume} {104}},\ \bibinfo {pages} {076803} (\bibinfo {year} {2010})}\BibitemShut {NoStop}%
\bibitem [{\citenamefont {Barkeshli}\ and\ \citenamefont {McGreevy}(2014)}]{barkeshli_prb_2014}%
  \BibitemOpen
  \bibfield  {author} {\bibinfo {author} {\bibfnamefont {M.}~\bibnamefont {Barkeshli}}\ and\ \bibinfo {author} {\bibfnamefont {J.}~\bibnamefont {McGreevy}},\ }\bibfield  {title} {\bibinfo {title} {Continuous transition between fractional quantum hall and superfluid states},\ }\href {https://doi.org/10.1103/PhysRevB.89.235116} {\bibfield  {journal} {\bibinfo  {journal} {Phys. Rev. B}\ }\textbf {\bibinfo {volume} {89}},\ \bibinfo {pages} {235116} (\bibinfo {year} {2014})}\BibitemShut {NoStop}%
\bibitem [{Note3()}]{Note3}%
  \BibitemOpen
  \bibinfo {note} {This is particularly clear in the parton picture, as $\langle \Phi \rangle \protect \ne 0$ allows us to identify $f$ and $c$, as the dynamical gauge field $a$ is Higgsed. The resulting LFL quasiparticle residue will be given by $Z\sim |\langle \Phi \rangle |^2$. One can also understand this phase by going to the dual vortex picture, \begin {equation} \protect \mathcal {L}_{total}=\protect \mathcal {L}_{FS}[f,-a]+\protect \frac {i}{2\pi }\protect \tilde {a}\wedge d(a+A)+\protect \mathcal {L}[v_{\Phi },\protect \tilde {a}]+\protect \cdots , \end {equation} where $\Phi $ carries flux under $\protect \tilde {a}$. When the boson vortex $v_{\Phi }$ is gapped, integrating out $\protect \tilde {a}$ leads to a mass term for $a_{\mu }$. The resulting Fermi surface state of $f$ is then exactly the ordinary metallic state of the physical electron $c$.}\BibitemShut {Stop}%
\bibitem [{\citenamefont {Song}\ and\ \citenamefont {Zhang}(2023)}]{song_scipost_2023}%
  \BibitemOpen
  \bibfield  {author} {\bibinfo {author} {\bibfnamefont {X.-Y.}\ \bibnamefont {Song}}\ and\ \bibinfo {author} {\bibfnamefont {Y.-H.}\ \bibnamefont {Zhang}},\ }\bibfield  {title} {\bibinfo {title} {{Deconfined criticalities and dualities between chiral spin liquid, topological superconductor and charge density wave Chern insulator}},\ }\href {https://doi.org/10.21468/SciPostPhys.15.5.215} {\bibfield  {journal} {\bibinfo  {journal} {SciPost Phys.}\ }\textbf {\bibinfo {volume} {15}},\ \bibinfo {pages} {215} (\bibinfo {year} {2023})}\BibitemShut {NoStop}%
\bibitem [{\citenamefont {Lieb}\ \emph {et~al.}(1961)\citenamefont {Lieb}, \citenamefont {Schultz},\ and\ \citenamefont {Mattis}}]{lieb_ann_1961}%
  \BibitemOpen
  \bibfield  {author} {\bibinfo {author} {\bibfnamefont {E.}~\bibnamefont {Lieb}}, \bibinfo {author} {\bibfnamefont {T.}~\bibnamefont {Schultz}},\ and\ \bibinfo {author} {\bibfnamefont {D.}~\bibnamefont {Mattis}},\ }\bibfield  {title} {\bibinfo {title} {Two soluble models of an antiferromagnetic chain},\ }\href {https://doi.org/10.1016/0003-4916(61)90115-4} {\bibfield  {journal} {\bibinfo  {journal} {Annals of Physics}\ }\textbf {\bibinfo {volume} {16}},\ \bibinfo {pages} {407} (\bibinfo {year} {1961})}\BibitemShut {NoStop}%
\bibitem [{\citenamefont {Oshikawa}(2000)}]{oshikawa_prl_2000}%
  \BibitemOpen
  \bibfield  {author} {\bibinfo {author} {\bibfnamefont {M.}~\bibnamefont {Oshikawa}},\ }\bibfield  {title} {\bibinfo {title} {Commensurability, excitation gap, and topology in quantum many-particle systems on a periodic lattice},\ }\href {https://doi.org/10.1103/PhysRevLett.84.1535} {\bibfield  {journal} {\bibinfo  {journal} {Phys. Rev. Lett.}\ }\textbf {\bibinfo {volume} {84}},\ \bibinfo {pages} {1535} (\bibinfo {year} {2000})}\BibitemShut {NoStop}%
\bibitem [{\citenamefont {Hastings}(2004)}]{hastings_2004_prb}%
  \BibitemOpen
  \bibfield  {author} {\bibinfo {author} {\bibfnamefont {M.~B.}\ \bibnamefont {Hastings}},\ }\bibfield  {title} {\bibinfo {title} {Lieb-schultz-mattis in higher dimensions},\ }\href {https://doi.org/10.1103/PhysRevB.69.104431} {\bibfield  {journal} {\bibinfo  {journal} {Phys. Rev. B}\ }\textbf {\bibinfo {volume} {69}},\ \bibinfo {pages} {104431} (\bibinfo {year} {2004})}\BibitemShut {NoStop}%
\bibitem [{Note4()}]{Note4}%
  \BibitemOpen
  \bibinfo {note} {As argued in \cite {song_prb_2024}, adding a flavor mass to the critical point leads to a trivial insulator. However, no such insulator of bosons is possible at $\nu =1/2$ unless translation symmetry is broken, leading us to conclude the flavor adjoint masses must break translation}\BibitemShut {NoStop}%
\bibitem [{\citenamefont {Mross}\ \emph {et~al.}(2010)\citenamefont {Mross}, \citenamefont {McGreevy}, \citenamefont {Liu},\ and\ \citenamefont {Senthil}}]{mross2010}%
  \BibitemOpen
  \bibfield  {author} {\bibinfo {author} {\bibfnamefont {D.~F.}\ \bibnamefont {Mross}}, \bibinfo {author} {\bibfnamefont {J.}~\bibnamefont {McGreevy}}, \bibinfo {author} {\bibfnamefont {H.}~\bibnamefont {Liu}},\ and\ \bibinfo {author} {\bibfnamefont {T.}~\bibnamefont {Senthil}},\ }\bibfield  {title} {\bibinfo {title} {Controlled expansion for certain non-{{Fermi-liquid}} metals},\ }\href {https://doi.org/10.1103/PhysRevB.82.045121} {\bibfield  {journal} {\bibinfo  {journal} {Physical Review B}\ }\textbf {\bibinfo {volume} {82}},\ \bibinfo {pages} {045121} (\bibinfo {year} {2010})}\BibitemShut {NoStop}%
\bibitem [{\citenamefont {Ye}\ \emph {et~al.}(2022)\citenamefont {Ye}, \citenamefont {Lee},\ and\ \citenamefont {Zou}}]{ye2022a}%
  \BibitemOpen
  \bibfield  {author} {\bibinfo {author} {\bibfnamefont {W.}~\bibnamefont {Ye}}, \bibinfo {author} {\bibfnamefont {S.-S.}\ \bibnamefont {Lee}},\ and\ \bibinfo {author} {\bibfnamefont {L.}~\bibnamefont {Zou}},\ }\bibfield  {title} {\bibinfo {title} {Ultraviolet-{{Infrared Mixing}} in {{Marginal Fermi Liquids}}},\ }\href {https://doi.org/10.1103/PhysRevLett.128.106402} {\bibfield  {journal} {\bibinfo  {journal} {Physical Review Letters}\ }\textbf {\bibinfo {volume} {128}},\ \bibinfo {pages} {106402} (\bibinfo {year} {2022})},\ \Eprint {https://arxiv.org/abs/2109.00004} {arXiv:2109.00004 [cond-mat]} \BibitemShut {NoStop}%
\bibitem [{\citenamefont {Morales-Dur\'an}\ \emph {et~al.}(2024)\citenamefont {Morales-Dur\'an}, \citenamefont {Wei}, \citenamefont {Shi},\ and\ \citenamefont {MacDonald}}]{duran_prl_2024}%
  \BibitemOpen
  \bibfield  {author} {\bibinfo {author} {\bibfnamefont {N.}~\bibnamefont {Morales-Dur\'an}}, \bibinfo {author} {\bibfnamefont {N.}~\bibnamefont {Wei}}, \bibinfo {author} {\bibfnamefont {J.}~\bibnamefont {Shi}},\ and\ \bibinfo {author} {\bibfnamefont {A.~H.}\ \bibnamefont {MacDonald}},\ }\bibfield  {title} {\bibinfo {title} {{Magic Angles and Fractional Chern Insulators in Twisted Homobilayer Transition Metal Dichalcogenides}},\ }\href {https://doi.org/10.1103/PhysRevLett.132.096602} {\bibfield  {journal} {\bibinfo  {journal} {Phys. Rev. Lett.}\ }\textbf {\bibinfo {volume} {132}},\ \bibinfo {pages} {096602} (\bibinfo {year} {2024})}\BibitemShut {NoStop}%
\bibitem [{\citenamefont {Paul}\ \emph {et~al.}(2023)\citenamefont {Paul}, \citenamefont {Zhang},\ and\ \citenamefont {Fu}}]{paul_sci_2023}%
  \BibitemOpen
  \bibfield  {author} {\bibinfo {author} {\bibfnamefont {N.}~\bibnamefont {Paul}}, \bibinfo {author} {\bibfnamefont {Y.}~\bibnamefont {Zhang}},\ and\ \bibinfo {author} {\bibfnamefont {L.}~\bibnamefont {Fu}},\ }\bibfield  {title} {\bibinfo {title} {Giant proximity exchange and flat {Chern band in 2D magnet-semiconductor heterostructures}},\ }\href {http://dx.doi.org/10.1126/sciadv.abn1401} {\bibfield  {journal} {\bibinfo  {journal} {Science Advances}\ }\textbf {\bibinfo {volume} {9}} (\bibinfo {year} {2023})}\BibitemShut {NoStop}%
\bibitem [{\citenamefont {Reddy}\ \emph {et~al.}(2023)\citenamefont {Reddy}, \citenamefont {Alsallom}, \citenamefont {Zhang}, \citenamefont {Devakul},\ and\ \citenamefont {Fu}}]{reddy_prb_2023}%
  \BibitemOpen
  \bibfield  {author} {\bibinfo {author} {\bibfnamefont {A.~P.}\ \bibnamefont {Reddy}}, \bibinfo {author} {\bibfnamefont {F.}~\bibnamefont {Alsallom}}, \bibinfo {author} {\bibfnamefont {Y.}~\bibnamefont {Zhang}}, \bibinfo {author} {\bibfnamefont {T.}~\bibnamefont {Devakul}},\ and\ \bibinfo {author} {\bibfnamefont {L.}~\bibnamefont {Fu}},\ }\bibfield  {title} {\bibinfo {title} {Fractional quantum anomalous {H}all states in twisted bilayer {M}o{T}e$_2$ and {WS}e$_2$},\ }\href {https://doi.org/10.1103/PhysRevB.108.085117} {\bibfield  {journal} {\bibinfo  {journal} {Phys. Rev. B}\ }\textbf {\bibinfo {volume} {108}},\ \bibinfo {pages} {085117} (\bibinfo {year} {2023})}\BibitemShut {NoStop}%
\bibitem [{\citenamefont {Greiter}\ \emph {et~al.}(1991)\citenamefont {Greiter}, \citenamefont {Wen},\ and\ \citenamefont {Wilczek}}]{greiter_prl_1991}%
  \BibitemOpen
  \bibfield  {author} {\bibinfo {author} {\bibfnamefont {M.}~\bibnamefont {Greiter}}, \bibinfo {author} {\bibfnamefont {X.-G.}\ \bibnamefont {Wen}},\ and\ \bibinfo {author} {\bibfnamefont {F.}~\bibnamefont {Wilczek}},\ }\bibfield  {title} {\bibinfo {title} {Paired hall state at half filling},\ }\href {https://doi.org/10.1103/PhysRevLett.66.3205} {\bibfield  {journal} {\bibinfo  {journal} {Phys. Rev. Lett.}\ }\textbf {\bibinfo {volume} {66}},\ \bibinfo {pages} {3205} (\bibinfo {year} {1991})}\BibitemShut {NoStop}%
\bibitem [{\citenamefont {Greiter}\ \emph {et~al.}(1992)\citenamefont {Greiter}, \citenamefont {Wen},\ and\ \citenamefont {Wilczek}}]{greiter_nucl_1992}%
  \BibitemOpen
  \bibfield  {author} {\bibinfo {author} {\bibfnamefont {M.}~\bibnamefont {Greiter}}, \bibinfo {author} {\bibfnamefont {X.}~\bibnamefont {Wen}},\ and\ \bibinfo {author} {\bibfnamefont {F.}~\bibnamefont {Wilczek}},\ }\bibfield  {title} {\bibinfo {title} {Paired hall states},\ }\href {https://doi.org/https://doi.org/10.1016/0550-3213(92)90401-V} {\bibfield  {journal} {\bibinfo  {journal} {Nuclear Physics B}\ }\textbf {\bibinfo {volume} {374}},\ \bibinfo {pages} {567} (\bibinfo {year} {1992})}\BibitemShut {NoStop}%
\bibitem [{Note5()}]{Note5}%
  \BibitemOpen
  \bibinfo {note} {This corresponds to $\epsilon =1$, in which the fixed point at $(\protect \overline {V},\protect \tilde {\alpha })=(0,0)$ splits into fixed points at $\protect \overline {V}=\pm \protect \sqrt {\epsilon /2}$. The fixed point $\protect \overline {V}=-\protect \sqrt {\epsilon /2}$ is unstable and $\protect \overline {V}=\protect \sqrt {\epsilon /2}$ is a stable fixed point, controlling the CFL phase. Therefore, with short range interactions, the CFL phase is still stable to pairing. Furthermore, the CFL in the proximity of the critical point is also stable to weak pairing, with the same reasoning as before. Even a distance away from the critical point into the CFL phase, there is an energy scale $\epsilon _*\sim m^{2\nu }$ below which the system is described by the CFL theory. Above $\epsilon _*$ the flow of $\protect \overline {V}$ tends to the attractor line $\protect \overline {V}=\protect \sqrt {\protect \tilde {\alpha }}$. Then, if $\protect \overline {V}(\epsilon _*)>-\protect \sqrt
  {\epsilon /2}$ at the crossover scale, no pairing will occur as we lower energy into the CFL regime.}\BibitemShut {Stop}%
\bibitem [{\citenamefont {Yankowitz}\ \emph {et~al.}(2019)\citenamefont {Yankowitz}, \citenamefont {Chen}, \citenamefont {Polshyn}, \citenamefont {Zhang}, \citenamefont {Watanabe}, \citenamefont {Taniguchi}, \citenamefont {Graf}, \citenamefont {Young},\ and\ \citenamefont {Dean}}]{yankowitz_science_2019}%
  \BibitemOpen
  \bibfield  {author} {\bibinfo {author} {\bibfnamefont {M.}~\bibnamefont {Yankowitz}}, \bibinfo {author} {\bibfnamefont {S.}~\bibnamefont {Chen}}, \bibinfo {author} {\bibfnamefont {H.}~\bibnamefont {Polshyn}}, \bibinfo {author} {\bibfnamefont {Y.}~\bibnamefont {Zhang}}, \bibinfo {author} {\bibfnamefont {K.}~\bibnamefont {Watanabe}}, \bibinfo {author} {\bibfnamefont {T.}~\bibnamefont {Taniguchi}}, \bibinfo {author} {\bibfnamefont {D.}~\bibnamefont {Graf}}, \bibinfo {author} {\bibfnamefont {A.~F.}\ \bibnamefont {Young}},\ and\ \bibinfo {author} {\bibfnamefont {C.~R.}\ \bibnamefont {Dean}},\ }\bibfield  {title} {\bibinfo {title} {Tuning superconductivity in twisted bilayer graphene},\ }\href {https://doi.org/10.1126/science.aav1910} {\bibfield  {journal} {\bibinfo  {journal} {Science}\ }\textbf {\bibinfo {volume} {363}},\ \bibinfo {pages} {1059} (\bibinfo {year} {2019})}\BibitemShut {NoStop}%
\bibitem [{\citenamefont {Wang}\ \emph {et~al.}(2025)\citenamefont {Wang}, \citenamefont {An}, \citenamefont {Ye}, \citenamefont {Wang}, \citenamefont {Mai}, \citenamefont {Zhao}, \citenamefont {Zhang}, \citenamefont {Peng}, \citenamefont {Watanabe}, \citenamefont {Taniguchi}, \citenamefont {Sun}, \citenamefont {Dai}, \citenamefont {Wang}, \citenamefont {Qin}, \citenamefont {Qiao},\ and\ \citenamefont {Zhang}}]{wang_prl_2025}%
  \BibitemOpen
  \bibfield  {author} {\bibinfo {author} {\bibfnamefont {Y.}~\bibnamefont {Wang}}, \bibinfo {author} {\bibfnamefont {J.}~\bibnamefont {An}}, \bibinfo {author} {\bibfnamefont {C.}~\bibnamefont {Ye}}, \bibinfo {author} {\bibfnamefont {X.}~\bibnamefont {Wang}}, \bibinfo {author} {\bibfnamefont {D.}~\bibnamefont {Mai}}, \bibinfo {author} {\bibfnamefont {H.}~\bibnamefont {Zhao}}, \bibinfo {author} {\bibfnamefont {Y.}~\bibnamefont {Zhang}}, \bibinfo {author} {\bibfnamefont {C.}~\bibnamefont {Peng}}, \bibinfo {author} {\bibfnamefont {K.}~\bibnamefont {Watanabe}}, \bibinfo {author} {\bibfnamefont {T.}~\bibnamefont {Taniguchi}}, \bibinfo {author} {\bibfnamefont {X.}~\bibnamefont {Sun}}, \bibinfo {author} {\bibfnamefont {R.}~\bibnamefont {Dai}}, \bibinfo {author} {\bibfnamefont {Z.}~\bibnamefont {Wang}}, \bibinfo {author} {\bibfnamefont {W.}~\bibnamefont {Qin}}, \bibinfo {author} {\bibfnamefont {Z.}~\bibnamefont {Qiao}},\ and\ \bibinfo {author} {\bibfnamefont {Z.}~\bibnamefont {Zhang}},\ }\bibfield  {title} {\bibinfo
  {title} {Pressure-driven moir\'e potential enhancement and tertiary gap opening in graphene/h-{BN} heterostructure},\ }\href {https://doi.org/10.1103/xs5j-hp3p} {\bibfield  {journal} {\bibinfo  {journal} {Phys. Rev. Lett.}\ }\textbf {\bibinfo {volume} {135}},\ \bibinfo {pages} {046303} (\bibinfo {year} {2025})}\BibitemShut {NoStop}%
\bibitem [{\citenamefont {Chen}\ \emph {et~al.}(1993)\citenamefont {Chen}, \citenamefont {Fisher},\ and\ \citenamefont {Wu}}]{chen_prb_1993}%
  \BibitemOpen
  \bibfield  {author} {\bibinfo {author} {\bibfnamefont {W.}~\bibnamefont {Chen}}, \bibinfo {author} {\bibfnamefont {M.~P.~A.}\ \bibnamefont {Fisher}},\ and\ \bibinfo {author} {\bibfnamefont {Y.-S.}\ \bibnamefont {Wu}},\ }\bibfield  {title} {\bibinfo {title} {Mott transition in an anyon gas},\ }\href {https://doi.org/10.1103/PhysRevB.48.13749} {\bibfield  {journal} {\bibinfo  {journal} {Phys. Rev. B}\ }\textbf {\bibinfo {volume} {48}},\ \bibinfo {pages} {13749} (\bibinfo {year} {1993})}\BibitemShut {NoStop}%
\bibitem [{\citenamefont {Senthil}(2008)}]{senthil_prb_2008}%
  \BibitemOpen
  \bibfield  {author} {\bibinfo {author} {\bibfnamefont {T.}~\bibnamefont {Senthil}},\ }\bibfield  {title} {\bibinfo {title} {Theory of a continuous {Mott} transition in two dimensions},\ }\href {https://doi.org/10.1103/PhysRevB.78.045109} {\bibfield  {journal} {\bibinfo  {journal} {Phys. Rev. B}\ }\textbf {\bibinfo {volume} {78}},\ \bibinfo {pages} {045109} (\bibinfo {year} {2008})}\BibitemShut {NoStop}%
\bibitem [{\citenamefont {Xu}\ \emph {et~al.}(2020)\citenamefont {Xu}, \citenamefont {Wu}, \citenamefont {Jian},\ and\ \citenamefont {Xu}}]{xu_prb_2020}%
  \BibitemOpen
  \bibfield  {author} {\bibinfo {author} {\bibfnamefont {Y.}~\bibnamefont {Xu}}, \bibinfo {author} {\bibfnamefont {X.-C.}\ \bibnamefont {Wu}}, \bibinfo {author} {\bibfnamefont {C.-M.}\ \bibnamefont {Jian}},\ and\ \bibinfo {author} {\bibfnamefont {C.}~\bibnamefont {Xu}},\ }\bibfield  {title} {\bibinfo {title} {Orbital order and possible non-{F}ermi liquid in moir\'e systems},\ }\href {https://doi.org/10.1103/PhysRevB.101.205426} {\bibfield  {journal} {\bibinfo  {journal} {Phys. Rev. B}\ }\textbf {\bibinfo {volume} {101}},\ \bibinfo {pages} {205426} (\bibinfo {year} {2020})}\BibitemShut {NoStop}%
\bibitem [{\citenamefont {Mesaros}\ \emph {et~al.}(2017)\citenamefont {Mesaros}, \citenamefont {Lawler},\ and\ \citenamefont {Kim}}]{mesaros_prb_2017}%
  \BibitemOpen
  \bibfield  {author} {\bibinfo {author} {\bibfnamefont {A.}~\bibnamefont {Mesaros}}, \bibinfo {author} {\bibfnamefont {M.~J.}\ \bibnamefont {Lawler}},\ and\ \bibinfo {author} {\bibfnamefont {E.-A.}\ \bibnamefont {Kim}},\ }\bibfield  {title} {\bibinfo {title} {Nematic fluctuations balancing the zoo of phases in half-filled quantum {H}all systems},\ }\href {https://doi.org/10.1103/PhysRevB.95.125127} {\bibfield  {journal} {\bibinfo  {journal} {Phys. Rev. B}\ }\textbf {\bibinfo {volume} {95}},\ \bibinfo {pages} {125127} (\bibinfo {year} {2017})}\BibitemShut {NoStop}%
\bibitem [{\citenamefont {Xu}\ \emph {et~al.}(2025{\natexlab{b}})\citenamefont {Xu}, \citenamefont {Sun}, \citenamefont {Li}, \citenamefont {Zheng}, \citenamefont {Xu}, \citenamefont {Gao}, \citenamefont {Jia}, \citenamefont {Watanabe}, \citenamefont {Taniguchi}, \citenamefont {Tong}, \citenamefont {Lu}, \citenamefont {Jia}, \citenamefont {Shi}, \citenamefont {Jiang}, \citenamefont {Zhang}, \citenamefont {Zhang}, \citenamefont {Lei}, \citenamefont {Liu},\ and\ \citenamefont {Li}}]{xu2025arxiv}%
  \BibitemOpen
  \bibfield  {author} {\bibinfo {author} {\bibfnamefont {F.}~\bibnamefont {Xu}}, \bibinfo {author} {\bibfnamefont {Z.}~\bibnamefont {Sun}}, \bibinfo {author} {\bibfnamefont {J.}~\bibnamefont {Li}}, \bibinfo {author} {\bibfnamefont {C.}~\bibnamefont {Zheng}}, \bibinfo {author} {\bibfnamefont {C.}~\bibnamefont {Xu}}, \bibinfo {author} {\bibfnamefont {J.}~\bibnamefont {Gao}}, \bibinfo {author} {\bibfnamefont {T.}~\bibnamefont {Jia}}, \bibinfo {author} {\bibfnamefont {K.}~\bibnamefont {Watanabe}}, \bibinfo {author} {\bibfnamefont {T.}~\bibnamefont {Taniguchi}}, \bibinfo {author} {\bibfnamefont {B.}~\bibnamefont {Tong}}, \bibinfo {author} {\bibfnamefont {L.}~\bibnamefont {Lu}}, \bibinfo {author} {\bibfnamefont {J.}~\bibnamefont {Jia}}, \bibinfo {author} {\bibfnamefont {Z.}~\bibnamefont {Shi}}, \bibinfo {author} {\bibfnamefont {S.}~\bibnamefont {Jiang}}, \bibinfo {author} {\bibfnamefont {Y.}~\bibnamefont {Zhang}}, \bibinfo {author} {\bibfnamefont {Y.}~\bibnamefont {Zhang}}, \bibinfo {author} {\bibfnamefont
  {S.}~\bibnamefont {Lei}}, \bibinfo {author} {\bibfnamefont {X.}~\bibnamefont {Liu}},\ and\ \bibinfo {author} {\bibfnamefont {T.}~\bibnamefont {Li}},\ }\href {https://arxiv.org/abs/2504.06972} {\bibinfo {title} {Signatures of unconventional superconductivity near reentrant and fractional quantum anomalous {H}all insulators}} (\bibinfo {year} {2025}{\natexlab{b}}),\ \Eprint {https://arxiv.org/abs/2504.06972} {arXiv:2504.06972 [cond-mat.mes-hall]} \BibitemShut {NoStop}%
\end{thebibliography}%
\end{document}